\documentclass{sn-jnl}

\usepackage{graphicx}%
\usepackage{multirow}%
\usepackage{amsmath,amssymb,amsfonts}%
\usepackage{amsthm}%
\usepackage{mathrsfs}%
\usepackage[title]{appendix}%
\usepackage{xcolor}%
\usepackage{textcomp}%
\usepackage{manyfoot}%
\usepackage{hhline}%
\usepackage{algorithm}%
\usepackage{algorithmicx}%
\usepackage{listings}%
\usepackage{array}
\usepackage{enumerate}

\usepackage{mathtools}
\usepackage{thm-restate}
\usepackage{tkz-euclide}
\usepackage{tikz}
\usepackage{tabularray}

\newcommand{\real} {\mathbb{R}}
\newcommand{\eps} {\varepsilon}
\newcommand{\cancel}[1] {}
\newcommand{\norm}[1] {\|#1\|}

\def\eps{\varepsilon}
\def\RR{{\mathbb R}}
\def\NN{{\mathbb N}}

\def\XX{{\mathbb X}}
\def\PP{{\mathbb P}}
\def\Warping{{\mathcal W}}

\renewcommand{\emph}[1]{\textit{\textbf{#1}}}

\newtheorem{theorem}{Theorem}
\newtheorem{lemma}[theorem]{Lemma}%
\newtheorem{corollary}[theorem]{Corollary}%
\newtheorem{definition}{Definition}%

\newcommand{\new}[1]{{\textcolor{black}{ #1}}}
\definecolor{greycolor}{rgb}{0, 0, 0}

\newcounter{predcounter}

\newenvironment{predicate}[1][]{\refstepcounter{predcounter}($\mathcal P_{\thepredcounter}$): #1 }

\def\eps{\varepsilon}
\def\RR{{\mathbb R}}
\def\NN{{\mathbb N}}

\def\XX{{\mathbb X}}
\def\Pred{{\mathcal P}}

\def\dim{{d}}

\def\RSpace{{\mathcal{R}}}
\def\PSpace{{\mathcal{P}}}
\renewcommand{\emph}[1]{\textit{\textbf{#1}}}

\DeclarePairedDelimiter\Hausd{d_{H}(}{)}
\DeclarePairedDelimiter\rHausd{d_{\overrightarrow{H}}(}{)}
\DeclarePairedDelimiter\Frechet{d_{F}(}{)}
\DeclarePairedDelimiter\dFrechet{d_{\mathit{dF}}(}{)}
\DeclarePairedDelimiter\wFrechet{d_{\mathit{wF}}(}{)}
\DeclarePairedDelimiter\DTW{d_{\mathit{DTW}}(}{)}
\def\Warping{{\mathcal W}}

\title{Solving Proximity Problems for Polygonal Shapes via Polynomials}

\author[1]{\fnm{Frederik} \sur{Brüning}}\email{bruening@uni-bonn.de}
\equalcont{These authors contributed equally to this work.}

\author[2]{\fnm{Siu-Wing} \sur{Cheng}}\email{scheng@cse.ust.hk}
\equalcont{These authors contributed equally to this work.}

\author*[1]{\fnm{Anne} \sur{Driemel}}\email{driemel@cs.uni-bonn.de}
\equalcont{These authors contributed equally to this work.}

\author[2]{\fnm{Haoqiang} \sur{Huang}}\email{haoqiang.huang@connect.ust.hk}
\equalcont{These authors contributed equally to this work.}

\affil[1]{\orgname{University of Bonn}, \orgaddress{\street{Friedrich-Hirzebruch-Alle 8}, \city{Bonn}, \postcode{53115}, \country{Germany}}}

\affil[2]{\orgname{HKUST}, \orgaddress{\city{Hong Kong}, \country{China}}}

\keywords{VC-Dimension, Fréchet distance, Hausdorff distance, Dynamic Time Warping}

\begin{document}

\abstract{
\unboldmath
  We study range spaces, where the ground set consists of either polygonal curves in $\mathbb{R}^d$ or polygonal regions in the plane that may contain holes and the ranges are balls defined by distance measures, such as the Hausdorff distance, the Fréchet distance and the dynamic time warping distance. These range spaces appear in various applications like classification, simplification, range searching, density estimation and clustering when the instances are trajectories, time series or polygons. We show for the Fréchet distance of polygonal curves and the Hausdorff distance of polygonal curves and planar polygonal regions that the VC-dimension is upper-bounded by $O(dk\log(km))$, where $k$ is the complexity of the center of a ball, $m$ is the complexity of each polygonal curve or region in the ground set, and $d$ is the ambient dimension. For $d \geq 4$ this bound is tight in each of the parameters $d,k$ and $m$ separately. For a variant of dynamic time warping distance of polygonal curves using squared distances, our analysis directly yields an upper-bound of $O(\min(dk^2\log(m),dkm\log(k)))$. We also obtain exact solutions that are hitherto unknown for simplification, range searching, nearest neighbor search, and distance oracle.
}

{

\maketitle

}

\section{Introduction}

The Vapnik–Chervonenkis dimension (VC-dimension) is a measure of complexity for range spaces that is named after Vladimir Vapnik and Alexey Chervonenkis, who introduced the concept in their seminal paper~\cite{VC71}. The VC-dimension of a range space can be used to determine sample bounds for various computational tasks. These include sample bounds on the test error of a classification model in statistical learning theory \cite{VC2000} or sample bounds for an $\varepsilon$-net  \cite{haussler1987eps} or an $(\eta,\varepsilon)$-approximation \cite{Har-PeledS11} in computational geometry. Sample bounds based on the VC-dimension have been successfully applied in the context of kernel density estimation \cite{Joshi11}, neural networks \cite{AB99, Karp95}, coresets \cite{buchin22, conradi2024socg, Feldman2020, Feldman11}, clustering \cite{Brüning_Akitaya_Chambers_Driemel_2023, bruning22}, object recognition \cite{LindenbaumBen943d, LindenbaumB94} and other data analysis tasks.

We study range spaces, where the ground set consists of polygonal curves or polygonal regions and the ranges consist of balls defined by the Hausdorff distance.
Previous to our work,  Driemel, Nusser, Philips and Psarros \cite{driemel2021vc} derived almost tight bounds on the VC-dimension in the setting of polygonal curves. At the heart of their approach lies the definition of a set of boolean functions ({predicates}) which are based on the inclusion and intersection of simple geometric objects.  The predicates depend on the vertices of a center curve and a radius that defines a metric ball as well as the vertices of a query curve. The predicates are chosen such that, based on their truth values, one can determine whether the query curve is contained in the respective ball. Their proof of the VC-dimension bound uses the worst-case number of operations needed to determine the truth values of each predicate.

In this paper, we develop an extended set of predicates that can decide the Hausdorff distance between polygonal regions with holes in the plane. We give an improved analysis for the VC-dimension that considers each predicate as a combination of sign values of polynomials. This approach does not use the computational complexity of the distance evaluation, but instead uses the underlying structure of the range space defined by a system of polynomials directly.  Our techniques extend to Fréchet distance and dynamic time warping distance (DTW). By the lower bounds in \cite{driemel2021vc}, this approach directly leads to tight bounds for $d \geq 4$ for polygonal curves.

There are many algorithmic challenges that arise from the analysis of polygonal curves and regions and involve the considered range spaces.  Range searching under the Fr\'echet distance was made by ACM SIGSPATIAL in 2017 a software challenge~\cite{10.1145/3231541.3231549}.  There has been extensive research on simplification~\cite{agarwal2005near,bereg2008simplifying,bringmann2019polyline,BuchinDGPR22,cheng2022curve,ConradiKPR24,douglas1973algorithms,filtser18,guibas1993approximating,hershberger1992speeding,hiroshi1988polygonal,van2019global,van2018optimal}, clustering~\cite{BuchinDGPR22,buchin2019approximating,buchin2021approximating,cheng2022curve,ConradiKPR24,driemel2016clustering,NT20}, and nearest neighbor search~\cite{CH2023,DP2020,driemel2021ann,driemel2017locality,driemel2019sublinear,filtser2020approximate,indyk2002approximate,mirzanezhad2020approximate} for polygonal curves and regions.  Range searching has also been studied~\cite{afshani2018complexity}. The ability to describe the underlying structure of our range spaces by a system of polynomials, enables us to tackle several of these problems with tools from algebraic geometry.   Algebraic geometric tools have hardly been used before.  Afshani and Driemel~\cite{afshani2018complexity} employed a semialgebraic range searching solution in $\real^2$, but we go much further to use arrangements of polynomials in higher dimensions.

\subsection{Preliminaries}

\subsubsection{Polygonal curves and regions}

Let $\XX^d_k$ denote the space of polygonal curves in $\RR^d$ of $k$ vertices.  Let $\PP_k$ denote the space of polygonal regions in $\RR^2$ of $k$ vertices. For any element $P$ in $\XX^d_k$ or $\PP_k$, we call $k$ the \emph{size} of $P$ and denote it by $|P|$.

We call a polygonal curve $P$ a \emph{closed curve} if its first and last vertices are equal. We call $P$ \emph{self-intersecting} if there are two edges $e$ and $e'$ of $P$ such that $e \cap e' \not= \emptyset$ and $e \cap e'$ is not a common endpoint of $e$ and $e'$.  

In the case that $P$ is a closed curve in $\RR^2$ which is not self-intersecting, we call the union of $P$ and its interior a \emph{simple polygonal region} $S$ (without holes). We denote with $\partial S$ the boundary of $S$, which is $P$. Given a simple polygonal region $S_0$ and a set of pairwise disjoint simple polygonal regions $S_1,\dots, S_h$ in the interior of $S_0$, we also consider the set $S=S_0 - (S_1\cup\dots\cup S_h)$ a polygonal region and we call $S_1,\dots,S_h$ the \emph{holes} of $S$.

\subsubsection{Distance measures}

Let $\norm{\cdot}$ denote the standard Euclidean norm. Let $X,Y\subseteq \RR^\dim$ for some $\dim\in\NN$. The \emph{directed Hausdorff distance} from $X$ to $Y$ is defined as
\[\rHausd{X,Y}=\sup_{x\in X} \inf_{y\in Y}\norm{x-y}\]
and the \emph{Hausdorff distance} between $X$ and $Y$ is defined as
\[\Hausd{X,Y}=\max\{\rHausd{X,Y},\rHausd{Y,X}\}.\]
If a set $X$ consists of a single point $p\in \RR^\dim$, we may write $p$ instead of $\{p\}$ to simplify the notation, e.g.   $\Hausd{p,Y}$ instead of $\Hausd{\{p\},Y}$. 

Given two curves in $X,Y \in \XX^d_m$ with vertex sets $V_X$ and $V_Y$, respectively, we define the \emph{discrete Hausdorff distance} between $X$ and $Y$ as
\[
d_{\mathit{dH}}(X,Y) = d_H(V_X,V_Y).
\]

For $m_1,m_2\in\NN$, each sequence
$(1, 1) = (i_1, j_1), (i_2, j_2), \dots , (i_M, j_M) = (m_1, m_2)$
such that $i_k -i_{k-1}$ and $j_k -j_{k-1}$ are either $0$ or $1$ for all $k$ is a \emph{warping path} from $(1, 1)$ to $(m_1, m_2)$. We denote with $\Warping_{m_1,m_2}$  the set of
all warping paths from $(1, 1)$ to $(m_1, m_2)$. For any two polygonal curves $P\in \XX_{m_1}^d$ with vertices $p_1,\dots,p_{m_1}$ and  $Q\in \XX_{m_2}^d$ with vertices $q_1,\dots,q_{m_2}$, we also write $\Warping_{P,Q} = \Warping_{m_1,m_2}$, and call elements of $\Warping_{P,Q}$
warping paths between $P$ and $Q$. The \emph{dynamic time warping distance} between $P$ and $Q$ is defined as
\[\DTW{P, Q} = \min_{w\in\Warping_{P,Q}}\sum_{(i,j)\in w}\norm{p_{i}-q_{j}}^2.\]
The above definition refers to the variant that is based on squared distances $\|p_i - q_j\|^2$ instead of just distances $\|p_i - q_j\|$.  This variant based on squared distances have also been used in both theoretical and applied works before~\cite{froese2022algorithmica,neamtu2018icde,tan2020datamining}.
A warping path  that attains the above minimum is also called an \emph{optimal warping path} between $P$ and $Q$. We denote with $\Warping_{m_1,m_2}^*\subset \Warping_{m_1,m_2}$ the set of warping paths $w$ such that there exist polygonal curves $P\in\XX_{m_1}^d$ and $Q\in\XX_{m_2}^d$ with this optimum warping path $w$. 

The \emph{discrete Fréchet distance} of two polygonal curves $P\in \XX_{m_1}^d$ with vertices $p_1,\dots,p_{m_1}$ and  $Q\in \XX_{m_2}^d$ with vertices $q_1,\dots,q_{m_2}$ is defined as
\[\dFrechet{P,Q}=\min_{w\in\Warping_{P,Q}}\max_{(i,j)\in w}\norm{p_{i}-q_{j}}.\]
In the continuous case, we define the \emph{Fréchet distance} of $P$ and $Q$ as
 \[ \Frechet{P,Q}= \inf_{\alpha,\beta:[0,1] \rightarrow [0,1]} \sup_{t \in [0,1]}
\| P(\alpha(t)) - Q(\beta(t)) \|, \] 

\noindent where $\alpha$ and $\beta$ range over all functions that are non-decreasing, surjective and continuous. We further define the \emph{weak Fréchet distance} of $P$ and $Q$ as 
\[ \wFrechet{P,Q}= \inf_{\alpha,\beta:[0,1] \rightarrow [0,1]} \sup_{t \in [0,1]}
\| P(\alpha(t)) - Q(\beta(t)) \|, \] 

\noindent where $\alpha$ and $\beta$ range over all continuous functions with $\alpha(0)=\beta(0)=0$ and $\alpha(1)=\beta(1)=1$.

\subsubsection{Range spaces and VC dimension}

Let $c$ be an element in $\XX^d_k$ or $\PP_k$.  Let $X$ be $\XX^d_m$ or $\PP_m$ for some $m$ depending on whether $c \in \XX^d_k$ or $\PP_k$, respectively.  Note that $m$ is not necessarily equal to $k$.  Let $\rho$ be a distance measure among $\{F, \mathit{dF}, \mathit{wF}, H, \mathit{dH}, \mathit{DTW}\}$.

We define the ball with radius $r$ and center $c$ on $X$ under the distance measure $d_\rho$ as 
\[B_{\rho,r}(c,X)=\{x\in X\;|\; d_\rho(x,c)\leq r\}.\]
We use $B_{r}(c)$ to denote a Euclidean ball of radius $r$ centered at a point $c \in \RR^d$.
%
We study the following range spaces with ground sets $\XX^d_m$ and $\PP_m$:
\begin{align*}
    \RSpace_{\rho,k} &= \bigl\{B_{\rho,r}(c,\XX^d_m)\;|\; r>0, c\in \XX^d_k \bigr\}, \\
    \PSpace_{\rho,k} &= \bigl\{B_{\rho,r}(c,\PP_m) \;|\; r>0, c \in \PP_k \bigr\}.
\end{align*}
We call $\XX^d_k$ and $\PP_k$ the \emph{center sets} of $\RSpace_{\rho,k}$ and $\PSpace_{\rho,k}$, respectively.

Let $X$ be a set. We call a set $\RSpace$ where any $R \in \RSpace$ is of the form $R \subseteq X$ a \emph{range space} with \emph{ground set} $X$. 
We say a subset $A \subseteq X$ is \emph{shattered} by $\RSpace$ if for any $A' \subseteq A$ there exists an $R \in \RSpace$ such that $A'=r \cap A$. The \emph{VC-dimension} of $\RSpace$ (denoted by $VCdim(\RSpace)$) is the maximal size of a set $A\subseteq X$ that is shattered by $\RSpace$. 

Let $\RSpace$ be a range space with ground set $X$.  Let $F$ be a class of real-valued functions defined on $\RR^\dim \times X$. For all $a\in\RR$, if $a \geq 0$, let $\mathrm{sgn}(a)=1$; otherwise, let $\mathrm{sgn}(a)=0$. We say that $\RSpace$ is a \emph{$t$-combination} of $\mathrm{sgn}(F)$ if there is a boolean function $g:\{0,1\}^t\rightarrow\{0,1\}$ and functions $f_1,\dots,f_t\in F$ such that for all $R \in\RSpace$ there is a parameter vector $y\in \RR^d$ such that
\[R=\bigl\{x\in X \;|\; g(\mathrm{sgn}(f_1(y,x)),\dots,\mathrm{sgn}(f_t(y,x)))=1 \bigr\}.\]

\subsubsection{Voronoi entities}

To bound the VC-dimension for $\PP_m$ under $d_H$, we will use properties of the  Voronoi diagram of a set of line segments in the plane. Let $X$ be a set of subsets (called sites) of $\RR^2$. The \emph{Voronoi region} $reg(A)$ of a site $A\in X$ consists of all points $p\in\RR^2$ for which $A$ is the closest among all sites in $X$, i.e.
\[\mathit{reg}(A)=\{p\in\RR^2\;|\;\rHausd{p,A}< \rHausd{p,B}\; \text{for all}\; B\in X\setminus\{A\}\}.\]
The \emph{Voronoi diagram} is the complement of the union of all regions $reg(A)$ with $A\in X$, so 
\[\mathit{vd}(X)=\RR^2 \setminus\bigcup_{A\in X} \mathit{reg}(A).\]
For any two sites $A,B\in X$, the \emph{bisector} of $A$ and $B$ is
\[\mathit{bisec}(A,B)=\{p\in\RR^2\;|\;\rHausd{p,A}= \rHausd{p,B}\}.\]
The \emph{Voronoi edge} of $A,B\in X$ is defined as \[\mathit{ve}(A,B)=\mathit{vd}(X)\cap \mathit{bisec}(A,B),\] and the \emph{Voronoi vertices} of $A,B,C\in X$ are defined as \[\mathit{vv}(A,B,C)=\mathit{vd}(X)\cap \mathit{bisec}(A,B)\cap \mathit{bisec}(B,C).\]

\subsection{Previous work}

\vspace{4pt}

\noindent {\bf VC dimension.}  Upper and lower bounds for the VC dimension of $\mathcal{R}_{\rho,k}$ have been established~\cite{driemel2021vc}. 
The upper bounds include an $O(dk\log(dkm))$ bound for $\mathcal{R}_{\mathit{dF},k}$, an $O(d^2k^2\log (dkm))$ bound for $\mathcal{R}_{H,k}$ and $\mathcal{R}_{F,k}$, and an $O(d^2k\log (dkm))$ bound for $\mathcal{R}_{\mathit{wF},k}$.  The lower bounds include an $\Omega\bigl(\max\{dk\log k, \log (dm)\}\bigr)$ bound for $\mathcal{R}_{\rho,k}$ for any $\rho \in \{H,\mathit{dH},F,\mathit{dF},\mathit{wF}\}$ and any $d \geq 4$, and an $\Omega(\max\{k,\log m\})$ bound for $\mathcal{R}_{\rho,k}$ for any $\rho \in \{H,\mathit{dH},F,\mathit{dF},\mathit{wF}\}$ in $\real^2$.

\vspace{9pt}

\noindent {\bf Simplification.}  
%
Given any polygonal curve $P \in \XX^d_m$ and $r > 0$, one problem is to find a curve $S \subset \RR^d$ with minimum size such that the distance between $P$ and $S$ is at most $r$.  Let $\kappa(P,r)$ denote this minimum size.  The most prominent heuristics are line simplifications~\cite{douglas1973algorithms,hershberger1992speeding,hiroshi1988polygonal}.  They are fast and effective in practice, but they offer no guarantee of the output quality for any of our considered distance measures.  We summarize the results with guarantees on the output quality in Table~\ref{tb:simplify}.

{\renewcommand{\arraystretch}{1.5}
\begin{table}
\centering
\begin{tabular}{|c|c|c|c|c|c|}
\hline
$\rho$ & $d_\rho(S,P) \leq $ & $|S| \leq $ & Time complexity & Ref. & Remark \\
\hline \hline
\multirow{5}*{$F$}  & $r$ & $\kappa(P,r)$ & $O(m^2\log^2 m)$ & \cite{guibas1993approximating} & $d \leq 2$  \\
\hhline{~-----}
 & $r$ & $\kappa(P,r/2)$ & $O(m\log m)$ & \cite{agarwal2005near} & $V_S \subseteq V_P$  \\
\hhline{~-----}
 & $(1+\eps)r$ & $2\kappa(P,r)-2$ & $O(\eps^{2-2d}m^2\log m\log\log m)$ & \cite{van2019global} & $\eps \in (0,1)$ \\
\hhline{~-----}
 & $r$ & $\kappa(P,r)$ & $O(\kappa(P,r) \cdot m^5)$ & \cite{van2018optimal} & $V_S \subseteq V_P$ \\
\hhline{~-----}
 & $(1+\eps)r$ & $(1+\alpha)\kappa(P,r)$ & $\tilde{O}\bigl(m^{O(1/\alpha)} \cdot \bigl(\frac{d}{\alpha\eps}\bigr)^{O(d/\alpha)}\bigr)$ & \cite{cheng2022curve} & $\alpha,\eps \in (0,1)$\\
\hline
\multirow{2}*{$F, \mathit{wF}$} & $r$ & $\kappa(P,r)$ & $O(m^3)$ & \cite{bringmann2019polyline,van2019global} & $V_S \subseteq V_P$ \\
\hhline{~-----}
 & $r$ & $\kappa(P,r)$ & 
\begin{tabular}{c}
$O(m)$ in $\real$, \\[-0.6em]
NP-hard in $\RR^d$ for $d \geq 2$ 
\end{tabular}
& \cite{van2019global} & $V_S \subset P$\\
\hline
\multirow{2}*{$\mathit{dF}$} & $r$ & $\kappa(P,r)$ & $O(m\log m)$ & \cite{bereg2008simplifying} & $d\leq 3$ \\
\hhline{~-----}
 & $r$ & $\kappa(P,r)$ & $O(m^2)$ & \cite{bereg2008simplifying} & 
$V_S \subseteq V_P$, $d\leq 3$ \\
\hline
$H$ & $r$ & $\kappa(P,r)$ & strongly NP-hard & \cite{van2018optimal} & $V_S \subseteq V_P$ \\
\hline
\end{tabular}
\caption{Global similarity results on computing simplification $S$ from $P$ for a given $r > 0$.  We use $V_S$ and $V_P$ to denote the vertex sets of $S$ and $P$, respectively.}
\label{tb:simplify}
\end{table}
}

Another simplification problem is that given an integer $k \geq 2$, compute a curve $S \in \XX^d_k$ that minimizes the distance to a curve $P \in \XX^d_m$.  For the Fr\'echet distance in $\real^d$, if the vertices of $S$ are restricted to be a subset of those of $P$, an algorithm was given~\cite{Godau1991ANM} to compute $S$ in $O(m^4\log m)$ time with the additional property that $d_F(S,P) \leq 7\,\mathrm{opt}$, where opt is the minimum possible Fr\'{e}chet distance if the vertices of $S$ are unrestricted. 
Subsequently, it was shown~\cite{agarwal2005near} how to compute in $O(m\log m)$ time a curve $S \in \XX_k^d$ such that the vertices of $S$ are a subset of those of $P$ and $d_F(S,P) \leq 4\,\mathrm{opt}$.  For discrete Fr\'{e}chet distance, the curve $S \in \XX^d_k$ that minimizes $d_{\mathit{dF}}(S,P)$ can be computed in $O(km\log m \log (m/k))$ time, or $O(m^3)$ time if the vertices of $S$ must be a subset of the vertices of $P$~\cite{bereg2008simplifying}.  For dynamic time warping, there are $(1+\varepsilon)$-approximation algorithms based on dynamic programming that consider all possible assignments of vertices of $P$ to vertices of $S$~\cite{BuchinDGPR22,ConradiKPR24}.

\vspace{9pt}

\noindent {\bf Range searching.}  Let $T$ be a set of $n$ curves in $\XX^d_m$ or regions in $\PP_m$.  Let $k \geq 2$ be a given integer.  The problem is to construct a data structure so that for any $r > 0$ and any query curve $P \in \XX^d_k$ or region $P \in \PP_k$ corresponding to the type of $T$, one can efficiently retrieve every $\tau \in T$ that has a distance of at most $r$ to $P$. 

Range search data structures and lower bounds have been developed for polygon curves under $d_{\mathit{dF}}$ in $\RR^d$ and under $d_F$ in $\RR^2$~\cite{afshani2018complexity}.  For $d_{\mathit{dF}}$ in $\RR^d$, an $O\bigl(n^{1-1/d}\log^{O(m)} n \cdot k^{O(d)}\bigr)$ query time was achieved using $O\bigl(n(\log\log n)^{m-1}\bigr)$ space.  The query time only accounts for the overhead spent to output the curves at a distance $r$ or less from the query curve in linear time.  For $d_F$ in $\RR^2$, the distance $r$ needs to be given for preprocessing and $k$ must be $\log^{O(1)} m$.  Under these two conditions, an $O\bigl(\sqrt{n}\log^{O(m^2)}n\bigr)$ query time was achieved using $O\bigl(n(\log\log n)^{O(m^2)}\bigr)$ space. As for lower bounds, Let $S(n)$ and $Q(n)$ be the space and query time of any range searching data structure for this problem, respectively.  For $d_{\mathit{dF}}$ and $d_F$, it was proved that $S(n) = \Omega\bigl(\frac{n^2}{Q(n)^2}\bigr) \cdot \bigl(\frac{\log(n/Q(n))}{\log\log n}\bigr)^{k-1}/2^{O(2^k)}$ and $S(n) = \Omega\bigl(\frac{n^2}{Q(n)^2}\bigr) \cdot \bigl(\frac{\log(n/Q(n))}{k^{3+o(1)}\log\log n}\bigr)^{k-1-o(1)}$. However, these lower bounds only hold for $d>1$.
Recently, better bounds were shown for one-dimensional curves. In~\cite{blankD24-rr}, rectangle-stabbing data structures were used to build a data structure with size $O(n \log^{k-1} n)$ and query time $O(\log^{k-1} n)$. Similarly, orthogonal range-searching data structures were used to obtain bounds depending on $m$ instead of $k$ by exploiting dual problems. 

The problem of counting the number of inclusion-maximal subcurves of a curve $\tau \in \mathbb{X}_m^d$ that are within a query Fr\'{e}chet distance $r$ from a query segment $\ell$ has been considered~\cite{DCG13}.  For any $s \in [m,m^2]$, an $\tilde{O}(m/\sqrt{s})$ query time was achieved using $\tilde{O}(s)$ space; however, the count may include some subcurves at Fr\'{e}chet distance up to $(2+3\sqrt{2})r$ from $\ell$.

\vspace{9pt}

\noindent {\bf Nearest neighbor.} Let $T$ be a set of $n$ curves in $\XX^d_m$ or regions in $\PP_m$.  Let $k \geq 2$ be a given integer.  The problem is constructing a data structure so that for any query $P \in \XX^d_k$ or $P \in \PP_k$ depending on the type of $T$,  its (approximate) nearest neighbor in $T$ can be retrieved efficiently.
No efficient exact solution is known for the considered distance measures.  

A related question is the $(\lambda,r)$-ANN query for any $\lambda > 1$ and any $r > 0$ given for preprocessing: given a curve $P \in \XX^d_k$ or region $P \in \PP_k$, either find $\tau \in T$ that is within a distance $\lambda r$ from $P$, or report that no  $\tau \in T$ is within a distance $r$ from $P$.  It was shown~\cite{HIM2012} that a $(\lambda,r)$-ANN data structure can be converted into a $\lambda$-ANN data structure.  Moreover, the storage and query time only increase by some polylogarithmic factors.

{\renewcommand{\arraystretch}{1.5}
\begin{table}
\begin{tabular}{|>{\centering\arraybackslash}p{0.8cm}
|>{\centering\arraybackslash}p{1.25cm}|>{\centering\arraybackslash}p{4cm}|>{\centering\arraybackslash}p{2.45cm}|c|p{1.45cm}|}
\hline
$\rho$ & $\lambda$ & Storage & Query time & Ref. & Remark \\
\hline
\hline
\multirow{14}*{$F$}  & $O(k)$ & $O(n\log n + mn)$ & $O(k\log n)$ & \cite{DP2020} & randomized, $d=1$ \\
\hhline{~-----}
 & $1+\eps$ & $n\cdot O(\frac{m}{k\varepsilon})^{k}$ & $O(2^kk)$ &  & \\
\hhline{~---~~}
 & $2+\eps$ & $n\cdot O(\frac{m}{k\varepsilon})^k$ & $O(k)$ &  & \\
\hhline{~---~~}
 & $2+\eps$ & $O(mn) + n \cdot O(\frac{1}{\varepsilon})^k$ & $O(2^kk)$ & \cite{BDNP2022} & $d=1$ \\
\hhline{~---~~}
 & $2+\eps$ & $O(mn)$ & $O(\frac{1}{\eps})^{k+2}$ &  &  \\
\hhline{~---~~}
 & $3+\eps$ & $O(mn) + n\cdot O(\frac{1}{\varepsilon})^k$ & $O(k)$ &  & \\ 
\hhline{~-----}
 & $1+\eps$ & $O(mn/\varepsilon)^{O(k)}\cdot\tilde{O}(k)$ &
$O(1/\varepsilon)^{O(k)} \cdot \tilde{O}(k)$ &  & $d\in\{2,3\}$ \\
\hhline{~---~-}
 & $1+\eps$ & 
$\tilde{O}\bigl(k(mnd^d/\varepsilon^d)^{O(k+1/\varepsilon^2)}\bigr)$ & $\tilde{O}\bigl(k(mn)^{0.5+\varepsilon}/\varepsilon^{d} +
k(d/\varepsilon)^{O(dk)}\bigr)$
& & $d \geq 4$ \\
\hhline{~---~-}
 & $3+\eps$ & 
$\tilde{O}(\frac{1}{\eps})^{O(dk)}(mn)^{O(k)}k$ & 
$\tilde{O}(k)$ & \cite{CH2023}  & $d \in \{2,3\}$ \\
\hhline{~---~-}
 & $3+\eps$ & $\tilde{O}(\frac{\sqrt{d}}{\eps})^{O(dk)}(mn)^{O(k)}k + O(\frac{\sqrt{d}}{\eps})^{O(d/\eps^2)} \cdot \tilde{O}((mn)^{O(1/\eps^2)})$ 
& $\tilde{O}(k(mn)^{0.5+\eps}/\eps^d)$ &  & $d \geq 4$ \\
\hhline{~-----}
 &  $1+\eps$ & $ (k \Phi/\eps)^{O(dk)} \cdot n $ & $ (k \Phi/\eps)^{O(dk)} \log n + (k \Phi/\eps)^{O(dk \log (1/\eps))}$ & \cite{ConradiDK24} & randomized preprocess, $\Phi$ is spread  \\
\hline
\hline
\multirow{9}*{$\mathit{dF}$} & $O(\log m + \log\log n)$ & $O(|X|^{\sqrt{m}}m^{2\sqrt{m}}n^2)$ & $O(m^{O(1)}\log n)$ & \cite{indyk2002approximate} & $X$:domain; $k \leq m$ \\
\hhline{~-----}
 & $O(m)$ & $O(n\log n + nm)$ & $O(m\log n)$ & \cite{driemel2017locality} & $k \leq m$\\
\hhline{~-----}
 & $1+\eps$ & $\bigl(2 + \frac{2}{\log m}\bigr)^{m^{O(1/\eps)}d\log(1/\eps)} \times \tilde{O}(n)$ &  $\tilde{O}(dm^{O(1/\eps)}) \times O(2^{4m}\log n)$ & \cite{emiris2020products} & randomized; $k \leq m$ \\
\hhline{~-----}
 & $1+\eps$ & $O(1/\eps)^{dk} \cdot n$ & $O(dk)$ & \multirow{3}*{\cite{filtser2020approximate}} & randomized preprocess \\
\hhline{~---~-}
 & $1+\eps$ & $O(1/\eps)^{dk} \cdot n$ & $O(dk\log\frac{nkd}{\eps})$ &  & \\
\hline
\hline
\multirow{6}*{$\mathit{DTW}$} & $O(m)$ & $O(n\log n + nm)$ & $O(m\log n)$ & \cite{driemel2017locality} & $k \leq m$\\
\hhline{~-----}
 & $1+\eps$ & $\tilde{O}(n) \cdot (1/\eps)^{O(dm)}$ & $\tilde{O}(d2^{4m}\log n)$ & \cite{emiris2020products} & randomized; $k \leq m$ \\
\hhline{~-----}
 & $1+\eps$ & $O(1/\eps)^{(d+1)m} \cdot n$ & $O(dm)$ & \cite{filtser2020approximate} & randomized preprocess; $k \leq m$ \\
\hline
\end{tabular}
\caption{$(\lambda,r)$-ANN results. There are $n$ input curves from $\XX^d_m$.  The query curve belongs to $\XX^d_k$.  By ``randomized'', we mean that the query succeeds probabilistically.  By ``randomized preprocess", we mean that the construction of the data structure is non-deterministic. }
\label{tb:nn}
\end{table}
}

Table~\ref{tb:nn} shows the $(\lambda,r)$-ANN data structures known for curves under several distance measures.  Several of the range searching data structures, such as ~\cite{afshani2018complexity, blankD24-rr}, can be adapted for the exact near-neighbor problem with little overhead. 

For the Hausdorff distance,  Farach-Colton and Indyk \cite{Farach-ColtonI99} obtain approximation results by embedding into $l_\infty$ and using known results for this target metric.
Bringman et al.~\cite{BDNP2022} proves some conditional lower bounds for $(\lambda,r)$-ANN.  For all $\eps, \eps' \in (0,1)$, one cannot achieve the following combinations of $(\lambda, \text{space}, \text{query time})$: $(2-\varepsilon, \text{poly}(n), O(n^{1-\varepsilon'}))$ in $\real$ when $1\ll k \ll \log n$ and $m = kn^{\Theta(1/k)}$,  $(3-\eps, \text{poly}(n), O(n^{1-\eps'}))$ in $\real$ when $k=m=\Theta(\log n)$, and $(3-\eps, \text{poly}(n), O(n^{1-\eps'}))$ in $\real^2$ when $1\ll k \ll \log n$ and $m = kn^{\Theta(1/k)}$.  

\vspace{9pt}

\noindent {\bf Distance oracle.} 
In some applications, given a curve $P \in \mathbb{X}_m^d$ and an integer $k \geq 2$, one wants to construct a \emph{distance oracle} so that for any curve $Q \in \mathbb{X}_k^d$, $d_F(Q,P)$ can be determined in $o(km)$ time.

In $\real^2$, a data structure of $O(m^2\log m)$ size was designed~\cite{DMO2017} such that for any \emph{horizontal} segment $\ell$ and any subcurve $P' \subseteq P \in \XX^2_d$, one can return $d_F(\ell,P')$ in $O(\log^2 m)$ time.  Recently, this result has been improved and extended in several ways in $\real^2$~\cite{buchin2022efficient}.  First, a data structure of $O(m\log m)$ size has been developed that answers a query for any horizontal segment in $O(\log m)$ time.  Second, there is a data structure of $O(m \log^2m)$ size that can answer the query for any horizontal segment and any subcurve of $P$ in $O(\log^3 m)$ time.  Third, queries for segments of arbitrary orientations can also be supported. 
There is a data structure that works for query curves in $\XX^2_k$ in a special case~\cite{GMMW2019}: for any $r > 0$ and any $Q \in \mathbb{X}_k^2$, if the edges in $Q$ and $P$ are suitably longer than $r$, one can decide whether $d_F(Q,P) \leq r$ in $O(k\log^2 m)$ time using $O(m\log m)$ space.

In $\mathbb{R}^d$, there are two approximate distance oracles for Fr\'{e}chet distance~\cite{DH2013}.  The first data structure works for any query segment $\ell$ and any subcurve $P'$ of $P$.  It uses $\tilde{O}(m\eps^{-2d})$ space and returns a $1+\eps$ approximation of $d_F(\ell,P')$ in $\tilde{O}(\eps^{-2})$ time.  The second data structure works for any 
query curve $Q \in \mathbb{X}_k^d$ and any subcurve $P'$ of $P$. 
It uses $O(m\log m)$ space and returns an $O(1)$-approximation of $d_F(Q,P')$ in $\tilde{O}(k^2)$ time.  

For lower bounds, it has been shown~\cite{bringmann2014walking} that, 
conditioned on SETH, it is impossible to construct an oracle in polynomial space and time that can return a 1.001 approximation of $d_F(Q,P)$ for any $Q \in \mathbb{X}_k^d$ in $O((km)^{1-\delta})$ time for any $\delta > 0$. For one-dimensional curves it is possible to break this barrier. In~\cite{blankD24-faster1d}, a distance oracle with preprocessing time and space in $O(m \log m)$ and query time in $O(k^2 \log^2 m)$.

For discrete Fr\'{e}chet distance, there are also exact and approximate distance oracles.  In~\cite{AFKR2024}, exact distance oracles have been designed for query curves in $\XX^2_k$ for $k \in [1,4]$ and input curves in $\XX^2_m$ for any $m \in \NN$.  For $k \in [1,3]$, the data structures use $O(m\log m)$ and answer a query in $O(\log^k m)$ time.  For $k = 4$, the data structure uses $O^*(m)$ space and answers a query in $O^*(\sqrt{m})$ time, where the $O^*(\cdot)$ notation hides subpolynomial factors.  In~\cite{FF2023}, $(1+\eps)$-approximate distance oracles were developed for query curves in $\XX^d_k$ and input curves in $\XX^d_m$.  It uses $O(1/\eps)^{kd} \cdot m\log(1/\eps)$ space and answers a query in $\tilde{O}(kd)$ time.  Queries on subcurves of the input curve can be supported at the expense of increasing the query time to $\tilde{O}(k^2d)^2$ and the space to $O(1/\eps)^{kd} \cdot m\log m \log(1/\eps)$.

\section{Our results}
We study problems for polygonal curves in $\RR^d$ and polygonal regions that may contain holes in $\RR^2$ with respect to several distance measures like the Hausdorff distance, the Fr\'echet distance and the dynamic time warping distance. For each distance measure $d_\rho$, we develop a set of polynomials of constant degree that characterize $d_\rho(P,Q)$ for any polygonal curves or regions $P$ and $Q$.  This allows us to construct an arrangement of the zero sets of polynomials such that each face of the arrangement encodes the exact solution for the problem in question.   

\vspace{9pt}

\noindent {\bf VC dimension.}  We derive new bounds on the VC-dimension for range spaces $\RSpace_{\rho,k}$ and $\mathcal{P}_{\rho,k}$. 
Recall that each element in $\mathcal{R}_{\rho,k}$ is the set of curves in $\XX^d_m$ that are within some distance $r$ under $d_\rho$ from some curve in $\XX^d_k$, and that each element in $\mathcal{P}_{\rho,k}$ is the set of polygonal regions in $\PP_m$, possibly with holes, that are within some distance $r$ under $d_\rho$ from some polygon region in $\PP_k$.

We prove that the VC dimension of $\mathcal{P}_{H,k}$ is bounded by $O(k\log(km))$.
Interestingly, this bound is independent of the number of holes. To the best of our knowledge this is the first non-trivial bound for polygonal regions under Hausdorff distance.
We remark that the construction of the lower bound of $\Omega(\max(k,\log(m)))$ for $d\geq 2$ in~\cite{driemel2021vc} for polygonal curves under Hausdorff distance can easily be generalized to a lower bound for the case of polygonal regions in the plane. To do so, we just have to replace each edge $e$ of a polygonal curve in their construction with a rectangle containing $e$ with suitable small width. This directly implies a bound of $\Omega(\max(k,\log(m)))$. Our upper bounds directly extend to unions of disjoint polygonal regions that may contain holes, where $k$ and $m$ denote the total number of edges.

Our techniques imply the same $O(dk\log(km))$ bound for the VC dimension of $\mathcal{R}_{\rho,k}$ for any $\rho \in \{H, \mathit{dH}, F, \mathit{dF}, \mathit{wF}\}$ and any dimension $d$.  This is an improvement over the previous VC dimension bounds of $O(dk\log(dkm))$ for $\mathcal{R}_{\mathit{dH},k}$ and $\mathcal{R}_{\mathit{dF},k}$, $O(d^2k^2\log(dkm))$ for $\mathcal{R}_{H,k}$ and $\mathcal{R}_{F,k}$, and $O(d^2k\log(dkm))$ for $\mathcal{R}_{\mathit{wF},k}$~\cite{driemel2021vc}.
By the known lower bound of $\Omega(\max(dk\log(k),\log(dm)))$ for $\mathcal{R}_{\rho,k}$ for any $d \geq 4$~\cite{driemel2021vc}, our new bounds are tight in each of the parameters $k,m$ and $d$ for each of the considered distance measures on polygonal curves. We also prove a new bound of $O\bigl(\min\bigl\{dk^2\log(m),dkm\log(k)\bigr\}\bigr)$ for the VC dimension of $\mathcal{R}_{\mathit{DTW},k}$.

\vspace{9pt}

\noindent {\bf Simplification.} Let $P$ be a given curve in $\XX^d_m$.  One simplification problem is to compute the curve $Q$ of the minimum complexity such that $d_\rho(P,Q)$ is not more than a given distance $r$.   We show that $Q$ can be computed in $O(km)^{O(dk)}$ time, where $k$ is the complexity of $Q$, if $\rho$ is any of the distance measure considered other than {\it DTW}.   For {\it DTW}, the running time is $\min \{O(m)^{O(dk^2)},O(k)^{O(dkm)}\}$. 

Another simplification problem is to compute a curve $Q$ of a given complexity $k \geq 2$ such that $d_\rho(P,Q)$ is minimized.  We show that $Q$ can be computed in $O(km)^{O(dk)}$ time if $\rho$ is any of the distance measure considered other than {\it DTW}.  For {\it DTW}, the running time is $\min \{O(m)^{O(dk^2)},O(k)^{O(dkm)}\}$.

Under the condition that the vertices of $Q$ are not restricted,
the above simplification algorithms are the first exact algorithms for all the distance measures considered other than discrete Fr\'{e}chet distance.

\vspace{9pt}

\noindent {\bf Range searching.}   Let $T$ be a set of $n$ polygonal curves in $\XX^d_m$ or  polygonal regions in $\PP_m$ that may contain holes.   Let $k \geq 2$ be a given integer and $d_\rho$ be one of the considered distance measures.  We present a data structure of $O(kmn)^{\mathrm{poly}(d,k)}$ size such that for any polygonal curve or region $Q$ with complexity $k$ and any $r > 0$, it returns every element $\tau \in T$ that satisfies $d_\rho(Q,\tau) \leq r$.  The query time is $O((dk)^{O(1)}\log (kmn))$ plus output size.

\vspace{9pt}

\noindent {\bf Nearest neighbor and distance oracle.}  Let $k \geq 2$ be a given integer and $d_\rho$ be one of the considered distance measures.  

Given a set $T$ of $n$ polygonal curves in $\XX^d_m$ or polygonal regions in $\PP_m$ that may contain holes, we can construct a nearest neighbor data structure of $O(kmn)^{\mathrm{poly}(d,k)}$ size such that for any polygonal curve or region $Q$ with complexity $k$, its nearest neighbor in $T$ under $d_\rho$ can be reported in $O((dk)^{O(1)}\log(kmn))$ time.  

Let $P$ be a curve in $\XX^d_m$ or region in $\PP_m$.  We can construct a distance oracle of $O(km)^{\mathrm{poly}(d,k)}$ size for $P$ such that for any polygonal curve or region $Q$ with complexity $k$, we can report $d_\rho(Q,P)$ in $O((dk)^{O(1)}\log(km))$ time.

\vspace{9pt}

In summary, we obtain improved bounds for  the VC dimension under the considered distance measures---by an order of magnitude in the case of $d_F$---that are close to the known lower bound, and we also obtain exact solutions for simplification, range searching, nearest neighbor search, and distance oracle. Exact solutions were not known for these problems in $\real^d$ for $d \geq 3$ except for curve simplification under discrete Fr\'{e}chet distance.  Exact solutions were not known for nearest neighbor search and distance oracle even in $\real^2$.  When $d$ and $k$ are in $O(1)$, our simplification algorithms run in polynomial time, and our data structures for the query problems use polynomial space and answer queries in logarithmic time.  Last but not least, the connection with arrangements of polynomials and algebraic geometry may offer new perspectives on designing algorithms and proving approximation results.

\section{Background results on algebraic geometry}\label{sec: pre}

We survey several algebraic geometric results that will be useful.  Given a set $\mathcal{P} = \{\rho_1,...,\rho_s\}$ of polynomials in $d$ real variables, a \emph{sign condition vector }$S$ for $\mathcal{P}$ is a vector in $\{-1, 0, +1\}^s$.  The point $\nu \in \mathbb{R}^{d}$ \emph{realizes} $S$ if $(\mathrm{sgn}(\rho_1(\nu)),...,\mathrm{sgn}(\rho_s(\nu))) = S$.  Let $\mathrm{sgn}(\mathcal{P}|_\nu)$ denote $(\mathrm{sgn}(\rho_1(\nu)),...,\mathrm{sgn}(\rho_s(\nu)))$.  The \emph{realization} of $S$ is the subset $\{\nu \in \real^d : \mathrm{sgn}(\mathcal{P}|_\nu) = S\}$.  For every $i \in [s]$, $\rho_i(\nu) = 0$ describes a hypersurface in $\real^d$.  The hypersurfaces $\rho_i(\nu) = 0$ for all $i \in [s]$ partition $\real^d$ into open cells of dimensions from 0 to $d$.  This set of cells together with the incidence relations among them form an \emph{arrangement} that we denote by $\mathscr{A}(\mathcal{P})$.  Each cell is a connected component of a realization of a sign condition vector for $\mathcal{P}$; the cells in $\mathscr{A}(\mathcal{P})$ represent all sign condition vectors that can be realized.  There are algorithms to construct a point in each cell of $\mathscr{A}(\mathcal{P})$ and optimize a function over a cell.

\vspace{9pt}

\begin{theorem}[\cite{Basu1995OnCA,basu2007algorithms,pollack1993number}]
\label{thm:arr}
Let $\mathcal{P}$ be a set of $s$ polynomials of $O(1)$ degree in $d$ variables.
\begin{enumerate}[{\em (i)}]
\item The number of cells in $\mathscr{A}(\+P)$ is $s^d \cdot O(1)^{d}$. 
\item A set $R$ of points can be computed in $s^{d+1} \cdot O(1)^{O(d)}$ time such that $R$ contains at least one point in each cell of $\mathscr{A}(\mathcal{P})$.  The sign condition vectors at these points are computed in the same time.
\item For a semialgebraic set $\mathcal{S}$ described using $\mathcal{P}$, the minimum over $\mathcal{S}$ of a polynomial of $O(1)$ degree in the same variables can be computed in $s^{2d+1}\cdot O(1)^{O(d)}$ time.  The point at the minimum is also returned.
\end{enumerate}
\end{theorem}	

\vspace{9pt}

\cancel{
	
	\begin{theorem}[Theorem 1~\cite{pollack1993number}]\label{thm: cell_num_bound}
		Given a set $\+P=\{P_1,...,P_s\}$ of $s$ polynomials in $d$ variables that have $O(1)$ degree, the number of cells in $\mathscr{A}(\+P)$ is $O(s^d)$. 
	\end{theorem}
	
	Besides upper bounding the number of cells, we are able to compute a point in every cell.
	
	\begin{theorem}[Theorem 2~\cite{Basu1995OnCA}]\label{thm: pt_for_cell}
		Given a set $\+P = \{P_1,...,P_s\}$ of polynomials in $d$ variables that have $O(1)$ degree, there is an algorithm that returns a set of points in $O(s^{d})$ time such that there are at least one point in each cell of $\mathscr{A}(\+P)$. Meanwhile, the algorithm also provides the sign condition realized by every point in the set. 
	\end{theorem} 
	
	The third theorem, which provides us with a way to find the infimum of a polynomial function in a cell of $\mathscr{A}(\+P)$, is also due to Basu, Pollack and Roy~\cite{basu2007algorithms}.
	
	\begin{theorem}[Algorithm 14.18~\cite{basu2007algorithms}]\label{thm: optimization}
		Given a set $\+P=\{P_1,...,P_s\}$ of polynomials in $d$ variables that have $O(1)$ degree, the minimum of a polynomial in the same variables that have $O(1)$ degree within the cell in $\mathscr{A}(\+P)$ can be computed in $O(s^{2d+1})$ time. Meanwhile, the vector that achieves the minimum value can also be generated.  
	\end{theorem}
	
}

We will need a \emph{point location} data structure for $\mathscr{A}(\mathcal{P})$ so that for any query point $\nu \in \real^d$, the cell in $\mathscr{A}(\mathcal{P})$ that contains $\nu$ can be reported quickly.  The point enclosure data structure proposed by Agarwal~et~al.~\cite{AAEZ2021} is applicable; however, the query time has a hidden factor that is exponential in the number of variables.  This is acceptable if there are $O(1)$ variables, but this may  not be so in our case.  Instead, we linearize the polynomials to hyperplanes in higher dimensions and use the point location solution by Ezra~et~al.~\cite{ezra2020decomposing}.

\vspace{9pt}

\begin{theorem}[\cite{ezra2020decomposing}]
	\label{thm:locate}
	Let $\mathcal{P}$ be a set of $s$ hyperplanes in $\real^d$.  For any $\eps > 0$, one can construct a data structure in $O(s^{2d\log d+O(d)})$ expected time and space such that any query point can be located in $\mathscr{A}(\mathcal{P})$ in $O(d^4\log s)$ time.
\end{theorem}

\vspace{9pt}

\cancel{
	\begin{theorem}[Theorem 2~\cite{MEISER1993286}]\label{thm: pt_location}
		Given a set $\+P=\{P_1,...,P_s\}$ of polynomials in $d$ variables that have degree 1, there is a data structure of space $O(s^{d+\varepsilon})$ that answers the point location query in the arrangement $\mathscr{A}(\+P)$ in $O(d^5\log s)$ time, where $\varepsilon$ is fixed non-negative value.
	\end{theorem}
}

The next result, which follows from the quantifier elimination result in~\cite[Algorithm~14.21]{basu2007algorithms}, gives the nature and complexity of an orthogonal projection of a semialgebraic set.

\vspace{9pt}

\begin{lemma}
	\label{lem:proj}
	Let $S$ be a semialgebraic set in $\real^d$ represented by $s$ polynomial inequalities and equalities in $d$ variables, each of degree at most $t$.  The orthogonal projection of $S$ in $\real^{d-1}$ along one of the axes is a semialgebraic set of $s^{2d}t^{O(d)}$ polynomial inequalities and equalities of degree at most $t^{O(1)}$.  The orthogonal projection can be computed in $s^{2d}t^{O(d)}$ time.
\end{lemma}

\vspace{9pt}

\cancel{
	
	Let $\+P=\{P_1,...,P_s\}\subset\mathbb{R}[x_1,...,x_h,y_1,...,y_\ell]$ be a set of polynomials in $h+\ell$ variable. A first order formula defined with respect to $\+P$ is 
	
	\begin{equation}\label{equation: first_order_formula}
		\Phi(y) = (\exists x_1,...,x_h)(\mathrm{sgn}(\+P(x,y))=S)
	\end{equation}
	where $x$ is a block of $h$ variables, $y$ is a block of $\ell$ free variables, and $S$ is a sign condition of $\+P$. A proposition known as a \emph{singly exponential quantifier elimination} shows that $\Phi(y)$ can be described by polynomial equalities and inequalities without the quantifier.
	
	\begin{theorem}\label{thm: quantifier_elimination}
		Let $\+P=\{P_1,...,P_s\}\subset\mathbb{R}[x_1,...,x_h,y_1,...,y_\ell]$ be a set of polynomials that have degree at most $t$. Given a first order formula $\Phi(y)$ of the form~\ref{equation: first_order_formula}, there is an algorithm running in $s^{(h+1)(\ell+1)}t^{O(hl)}$ time to return an equivalent quantifier-free formula
		
		\begin{equation}
			\Psi(y)=\bigvee_{i=1}^I\bigwedge_{j=1}^{J_i}\bigl(\bigvee_{n=1}^{N_{i,j}}\mathrm{sgn}(P_{ijn}(y))=b_{ijn}\bigr),
		\end{equation}
		where $P_{ijn}(y)$ is polynomial in $y$ that has degree bounded by $t^{O(h)}$, $b_{ijn}\in \{-1,0,+1\}$, and 
		
		\begin{align*}
			I&\le s^{(h+1)(\ell+1)}t^{O(hl)},\\
			J_i&\le s^{h+1}t^{O(h)},\\
			N_{i,j}&\le t^{O(h)}.
		\end{align*}
	\end{theorem}
	
	By quantifier elimination, we can project a cell specified by $\+P$ to some cells in the lower dimension.
}


We use the above described techniques from algebraic geometry throughout the paper. 
We will also use the following well-known theorem~\cite{AB99} which goes back to Goldberg and Jerrum~\cite{GoldbergJ93} and, independently, Ben-David and Lindenbaum~\cite{Ben-DavidL93}. 
The proof consists of expressing the range spaces as combinations of sign values of polynomials $\mathcal{P}$ and then bounding the VC-dimension based on the number of cells in $\mathscr{A}(\+P)$. 

\vspace{9pt}

\begin{restatable}[\cite{AB99}, Theorem 8.3]{theorem}{vcsimple}\label{thm:vcsimpl}
Let $F$ be a class of functions mapping from $\RR^\dim \times X$ to $\RR$ so 
that, for all $x\in X$ and $f\in F$ the function $y\rightarrow f(y,x)$ is a polynomial on $\RR^\dim$ of degree no more than $l$. Suppose that $\RSpace$ is a $t$-combination of $\mathrm{sgn}(F)$. Then we have 
\[
\mathit{VCdim}(\RSpace)\leq 2\dim\log_2(12tl).\]
\end{restatable}

\section{VC-dimension: Discrete distance measure}\label{sec:discrete}

\begin{theorem}\label{thm:disc:hausd}
$\mathit{VCdim}(\RSpace_{{\mathit{d}H},k})\leq 2(\dim k +1)\log_2(24 mk)$
\end{theorem}
\begin{proof}
Take a curve $P \in \XX^d_m$ with vertices $p_1,\dots,p_m$ and a curve $Q\in \XX^d_k$ with vertices $q_1,\dots,q_k$. The Hausdorff distance between two point sets is uniquely defined by the distances of the points of the two sets. The truth value of $d_{\mathit{dH}}{P,Q}\leq r$ can therefore be determined given the truth values of $\norm{p-q}^2\leq r^2$ for all pairs $(p,q)\in \{p_1,\dots,p_m\}\times\{q_1,\dots,q_k\}$. We can write the points $p,q\in\RR^d$ as tuples of their coordinates with $p=(p_{(1)},\dots,p_{(d)})$ and $q=(q_{(1)},\dots,q_{(d)})$. Then $\norm{p-q}^2\leq r^2$ is equivalent to
    \[r^2-\sum_{i=1}^\dim (p_{(i)}-q_{(i)})^2 \geq 0.\]
The term $r^2-\sum_{i=1}^\dim (p_{(i)}-q_{(i)})^2 $ is a polynomial of degree $2$ in all its variables. So the truth value of $\norm{p-q}^2\leq r^2$ can be determined by the sign value of a polynomial of degree 2. There are in total $mk$ possible choices for the pair $(p,q)$. Let $y\in \RR^{dk+1}$ be the vector consisting of all coordinates of the vertices $q_1,\dots,q_k$ and the radius $r$. Then $\RSpace_{{dH},k}$ is a $(mk)$-combination of $\mathrm{sgn}(F)$ where $F$ is a class of functions mapping from $\RR^{\dim k+1} \times \RR^{dm}$ to $\RR$ so 
that, for all $P\in \XX^d_m$ and $f\in F$ the function $y\rightarrow f(y,P)$ is a polynomial on $\RR^\dim$ of degree no more than $2$. The VC-dimension bound follows directly by applying Theorem~\ref{thm:vcsimpl}.
\end{proof}

\vspace{4pt}

The VC-dimension of $\mathcal{R}_{\mathit{dF},k}$ can be analyzed in the same way.

\vspace{9pt}

\begin{theorem}\label{thm:disc:frech}
$\mathit{VCdim}(\RSpace_{{\mathit{d}F},k})\leq 2(\dim k +1)\log_2(24 mk)$
\end{theorem}
\begin{proof}
    The proof is analogous to the proof of Theorem~\ref{thm:disc:hausd} given the fact that the discrete Fréchet distance between two polygonal curves is uniquely defined by the distances of the vertices of the two curves.
\end{proof}

\vspace{4pt}

\begin{theorem}\label{thm:disc:dtw}
$\mathit{VCdim}(\RSpace_{\mathit{DTW},k})
\leq 2(dk+1)\cdot\min\{(k-1)\log_2 m, (m-1)\log)_2k\} = O(\min\{dk^2\log(m),dkm\log(k)\})$
\end{theorem}
\begin{proof}
Take a curve $P \in \XX^d_m$ with vertices $p_1,\dots,p_m$ and a curve $Q\in \XX^d_k$ with vertices $q_1,\dots,q_k$. 
The truth value of $\DTW{P,Q}\leq r$ can be determined by the truth values of
$\sum_{(i,j)\in w}\norm{p_{i}-q_{j}}^2\leq r$ for all $w\in \mathcal{W}_{P,Q}$. This is equivalent to 
\[
\forall \, w \in \mathcal{W}_{P,Q}, \quad 
r-\sum_{(i,j)\in w}\sum_{t=1}^d (p_{i,t}-q_{j,t})^2\geq 0
\]
for which the left side is a polynomial of degree $2$ in all its variables. We get $\lvert\mathcal{W}_{P,Q}\rvert\leq \binom{m+k-2}{m-1} \leq \min \{ m^{k-1} , k^{m-1} \}$ by counting all possible optimal warping paths. Let $y\in \RR^{dk+1}$ be the vector consisting of all coordinates of the vertices $q_1,\dots,q_k$ and the radius $r$. Then $\RSpace_{{DTW},k}$ is a $\min \{ m^{k-1} , k^{m-1} \}$-combination of $\mathrm{sgn}(F)$ where $F$ is a class of functions mapping from $\RR^{\dim k+1} \times \RR^{dm}$ to $\RR$ so 
that, for all $P\in \XX^d_m$ and $f\in F$ the function $y\rightarrow f(y,P)$ is a polynomial on $\RR^\dim$ of constant degree. The VC-dimension bound follows directly by the application of Theorem~\ref{thm:vcsimpl}.
\end{proof}

\section{Characterizing distance queries with predicates}\label{sec:general:approach}
To bound the VC-dimension of range spaces of the form $\RSpace_{\rho,k}$ in the continuous setting  we define geometric predicates for distance queries with $d_{\rho}$. These predicates can for example consist of checking the distances of geometric objects or checking if some geometric intersections exist. They have to be chosen in a way that the query can be decided based on their truth values. We will show that our predicates can be viewed as combinations of simple predicates. 

\vspace{9pt}

\begin{definition}
    Let $F$ be a class of functions mapping from $\RR^{\dim m} \times \RR^{dk+1}$ to $\RR$ so 
that, for all $f\in F$, the function $(x,y)\rightarrow f(x,y)$ is a polynomial of constant degree. Let $\Pred$ be a function from $\RR^{\dim m} \times \RR^{dk+1}$ to $\{0,1\}$. We say that the predicate $\Pred$ is  \emph{simple} if $\Pred$ is a $t$-combination of $\mathrm{sgn}(F)$ with $t\in O(1)$.   We further say that an inequality is simple if its truth value is equivalent to a simple predicate. 
\end{definition}

\vspace{9pt}

In our proof of the VC-dimension bounds we will use the following corollary to Theorem~\ref{thm:vcsimpl}.

\vspace{9pt}

\begin{corollary}\label{cor:vc}
    Suppose that for a given $d_\rho$ there exists a polynomial $p(k,m)$ such that for any $k,m\in\NN$, the range space $\RSpace_{\rho,k}$ with ground set $\XX^d_m$ is a $p(k,m)$-combination of simple predicates. Then $\mathit{VCdim}(\RSpace_{\rho,k})$ is in
$O(dk\log(km))$.
\end{corollary}

\subsection{Encoding of the input}
To state the predicates, we introduce additional notation. Following~\cite{driemel2021vc}, we define the following \emph{basic geometric objects}.
Let $s,t\in \RR^\dim$ be two points, and let $r \in\RR_+$ be the radius. We denote with $\ell(\overline{st})$ the line supporting $\overline{st}$.  We define the stadium, cylinder and capped cylinder centered at $\overline{st}$ with radius $r$:
\begin{align*}
\text{Stadium:}~D_r(\overline{st})&=\{x\in \RR^\dim\;|\; \exists p \in \overline{st}, \norm{p-x}\leq r\}, \\
\text{Cylinder:}~C_r(\overline{st})&=\{x\in \RR^\dim\;|\;\exists p \in \ell(\overline{st}), \norm{p-x}\leq r\}, \\
\text{Capped cylinder:}~R_r(\overline{st})&=\{p+u\in \RR^\dim\;|\; p\in \overline{st}, \norm{u}\leq r, \langle t-s,u\rangle=0\}.
\end{align*}
We define the hyperplane through $s$ orthogonal to $\overline{st}$ as 
\[
P(\overline{st})=\{x\in \RR^\dim\;|\;\langle x-s,s-t\rangle=0\}.
\]
Let $e_1$ and $e_2$ be two edges. We define the double stadium of the  edges $e_1$ and $e_2$ with radius $r$:
\[
\text{Double stadium:} \quad D_{r,2}(e_1,e_2)=D_r(e_1)\cap D_r(e_2). 
\]
Let $p=(p_1,p_2)\in\RR^2$. We define the horizontal ray that shoots from $p$ to the right:
\[
\mathit{hr}(p)=\{(x_1,x_2)\in\RR^2\;|\; x_1\geq p_1, x_2= p_2\}.
\]
For two  polygonal curves $P\in \XX^d_m$ and $Q\in\XX^d_k$ and a radius $r$, each predicate is a function mapping from $\RR^{\dim m} \times \RR^{dk+1}$ to $\{0,1\}$ that receives the input $(P,Q,r)$. For two polygonal regions $P \in \PP_m$ and $Q \in \PP_k$ that may contain holes, we map from $\RR^{3m} \times \RR^{3k+1}$ since we add a label that associates each vertex with its boundary component. Other encodings are possible but would only influence the constant in the VC-dimension bound.

\subsection{Polygonal curves}
\label{sec:curves:Pred}

Take a curve $P\in \XX_m^d$ with vertices $p_1,\dots,p_m$ and  a curve $Q\in \XX_k^d$ with vertices $q_1,\dots,q_k$. Let $r$ be a positive value.  The Hausdorff distance query $\Hausd{P,Q}\leq r$ is uniquely determined by the following predicates~\cite{driemel2021vc}: 
\begin{itemize}

\item \begin{predicate}
     Given an edge of $P$, $\overline{p_j p_{j+1}}$, and a vertex $q_i$ of $Q$, this predicate returns true if and only if there exists a point $p\in \overline{p_j p_{j+1}}$, such that $\norm{p-q_i}\leq r$.\label{hpc1}
\end{predicate}

\item \begin{predicate}
Given an edge of $Q$, $\overline{q_i q_{i+1}}$, and a vertex $p_j$ of $P$, this predicate returns true if and only if there exists a point $q\in \overline{q_i q_{i+1}}$, such that $\norm{q-p_j}\leq r$.\label{hpc2}
\end{predicate} 

\item  \begin{predicate}
Given an edge of $Q$, $\overline{q_i q_{i+1}}$, and two edges $e_1$ and $e_2$ of $P$, this predicate is equal to $\ell(\overline{q_i q_{i+1}})\cap D_{r,2}(e_1,e_2)\neq \emptyset$.\label{hpc3}
\end{predicate} 

\item \begin{predicate}
Given an edge of $P$, $\overline{p_j p_{j+1}}$, and two edges $e_1$ and $e_2$ of $Q$, this predicate is equal to $\ell(\overline{p_j p_{j+1}})\cap D_{r,2}(e_1,e_2)\neq \emptyset$.\label{hpc4}
\end{predicate} 

\end{itemize}

\begin{figure}\centering\includegraphics[width=0.92\textwidth]{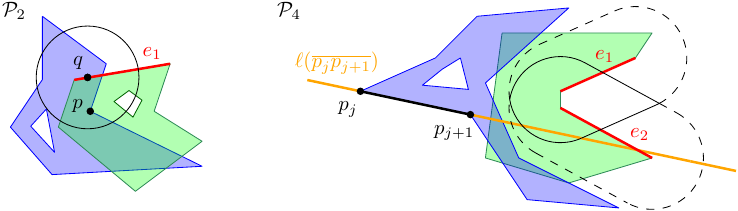}
    \caption{Illustration of the predicates $\Pred_{\ref{hpc2}}, \Pred_{\ref{hpc4}}$ :  In both depicted cases the corresponding predicates are true. Figure is best viewed in color. }
    \label{fig:polygon:predicates24}
\end{figure}

Examples for the predicates $\Pred_{\ref{hpc2}}$ and $\Pred_{\ref{hpc4}}$ for polygonal regions are depicted in Figure~\ref{fig:polygon:predicates24}. 

\vspace{9pt}

\begin{lemma}[Lemma 7.1, \cite{driemel2021vc}]\label{lem:hausd:pred}
For any two polygonal curves $P$ and $Q$, given the truth values of all predicates of the types $\Pred_{\ref{hpc1}}$, $\Pred_{\ref{hpc2}}$, $\Pred_{\ref{hpc3}}$, and $\Pred_{\ref{hpc4}}$, one can determine whether $\Hausd{P,Q}\leq r$. No additional knowledge of $P$ or $Q$ besides the truth values of these predicates is needed.
\end{lemma}

\vspace{9pt}

The Fr\'echet distance query $\Frechet{P,Q}\leq r$ is uniquely determined by the predicates $(\Pred_{\ref{hpc1}})$, $(\Pred_{\ref{hpc2}})$ and the following predicates~\cite{driemel2021vc}: 

\begin{itemize}

\item \begin{predicate}
This predicate returns true if and only if $\norm{p_1-q_1}\leq r$.\label{fpc1}
\end{predicate}

\item \begin{predicate}
This predicate returns true if and only if $\norm{p_m-q_k}\leq r$.\label{fpc2}
\end{predicate} 

\item \begin{predicate}
Given two vertices $p_j$ and $p_t$ of $P$, where $j < t$, and an edge $\overline{q_i q_{i+1}}$ of $Q$,  this predicate returns true if there exist two points $a_1$ and $a_2$ on $\ell(\overline{q_iq_{i+1}})$ such that $a_2-a_1$ has the same orientation as $q_{i+1}-q_i$, $\norm{a_1-p_j}\leq r$, and $\norm{a_2-p_t}\leq r$.\label{fpc3}
\end{predicate} 

\item \begin{predicate}
Given two vertices of $q_i$ and $q_t$ of $Q$, where $i < t$, and an edge $\overline{p_j p_{j+1}}$ of $P$, this predicate returns true if there exist two points $a_1$ and $a_2$ on the $\ell(\overline{p_jp_{j+1}})$ such that $a_2-a_1$ has the same orientation as $p_{j+1}-p_j$, $\norm{a_1-q_i}\leq r$, and $\norm{a_2-q_t}\leq r$.
\label{fpc4}
\end{predicate} 
\end{itemize}

\vspace{6pt}

\begin{lemma}[Lemma 7.1, \cite{AD18}]\label{lem:frechet:pred}
For any two polygonal curves $P$ and $Q$, given the truth values of all predicates of the types $\Pred_{\ref{hpc1}}$, $\Pred_{\ref{hpc2}}$, $\Pred_{\ref{fpc1}}$, $\Pred_{\ref{fpc2}}$, $\Pred_{\ref{fpc3}}$, and $\Pred_{\ref{fpc4}}$, one can determine whether $\Frechet{P,Q}\leq r$.  No additional knowledge of $P$ or $Q$ besides the truth values of these predicates is needed.
\end{lemma}

\vspace{9pt}

\begin{lemma}[Lemma 8.2, \cite{driemel2021vc}]\label{lem:wfrechet:pred}
For any two polygonal curves $P$ and $Q$, given the truth values of all predicates of the types $\Pred_{\ref{hpc1}}$, $\Pred_{\ref{hpc2}}$, $\Pred_{\ref{fpc1}}$, and $\Pred_{\ref{fpc2}}$, one can determine whether $\wFrechet{P,Q}\leq r$.  No additional knowledge of $P$ or $Q$ besides the truth values of these predicates is needed.
\end{lemma}

\subsection{Polygonal regions}\label{sec:Hausd:Pred:Pol}
In the case of polygonal regions that may contain holes, we define some predicates based on the Voronoi vertices induced by the boundary edges of the region.  Since degenerate situations can occur if Voronoi sites intersect, we restrict the predicates to the subset of the Voronoi vertices relevant to our analysis.

\subsubsection{Relevant Voronoi vertices}
\begin{figure} \centering\includegraphics[width=0.5\textwidth]{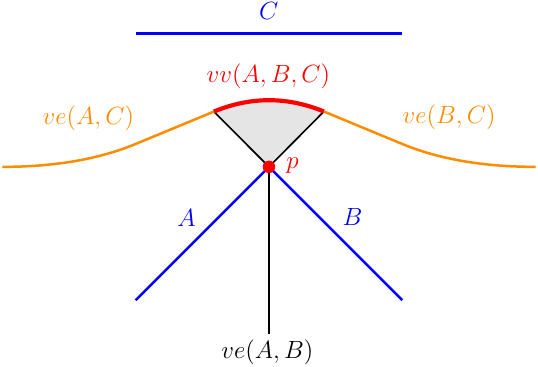}
    \caption{Degenerate case: $vv(A,B,C)$ consist of a whole arc and $ve(A,B)$ contains a region. Figure is best viewed in color.}
    \label{fig:deg_vert}
\end{figure}
If $A, B$ and $C$ are line segments and $A$ and $B$ intersect in a point $p$, it can happen  that there are Voronoi vertices in $vv(A,B,C)$ for which the closest point in $A$ and $B$ is $p$. This may result in degenerate cases where a whole arc of the Voronoi diagram consists of Voronoi vertices and a region is part of a Voronoi edge (see Figure~\ref{fig:deg_vert}). For our distance queries, we are only interested in  extreme points of the distance to the sites. These are Voronoi vertices that are not degenerate. We define the \emph{relevant Voronoi vertices} as the Voronoi vertices for which the distance of the vertices to the sites is realized by at least three distinct points, i.e. 
\begin{eqnarray*}
&&\mathit{rvv}(A,B,C)\\
&=&\left\{p\in \mathit{vv}(A,B,C)\ \middle\vert
\begin{array}{l}
 \exists \,\, \text{distinct} \,\,  x \in A, y \in B, z \in C \,\, \text{s.t.} \\ \rHausd{p,x}= \rHausd{p,y}=  \rHausd{p,z}=\rHausd{p,A\cup B\cup C}
\end{array}\right\}.
\end{eqnarray*}
Additionally, we introduce the notion of Voronoi-vertex-candidates. Let $a=\overline{a_1 a_2}, b=\overline{b_1 b_2}$ and $c=\overline{c_1 c_2}$ be edges of a polygonal region that may contain holes. Consider their vertices and supporting lines $A=\{a_1,a_2,\ell(a)\}$, $B=\{b_1,b_2,\ell(b)\}$ and $C=\{c_1,c_2,\ell(c)\}$. Let $X\in A$, $Y\in B$ and $Z\in C$. Define:
\begin{quote}
\begin{tabular}{l}
$V_0(X,Y,Z)=\emptyset$, if $X$, $Y$, or $Z$ is a subset of one of the other two. \\[.2em]
$V_0(X,Y,Z) = \left\{v\in\RR^2\;|\; \rHausd{v,X}=\rHausd{v,Y}=\rHausd{v,Z}\right\}$ otherwise.
\end{tabular}
\end{quote}
Note that $V_0(X,Y,Z)$ is the set of points with the same distance to $X$, $Y$, and $Z$.
The \emph{set of Voronoi-vertex-candidates $V(a,b,c)$} of the line segments $a, b$ and $c$ is defined as the union over all points that have the same distance to fixed elements of $A$, $B$ and $C$, i.e.
\[V(a,b,c)=\bigcup_{X\in A, Y\in B, Z\in C} V_0(X,Y,Z).\]
Note that the Voronoi-vertex-candidates $V(a,b,c)$ contain all relevant Voronoi vertices $rvv(a,b,c)$ because a relevant Voronoi vertex $v$ has the same distance to all three edges and for each edge, the distance to $v$ is either realized at one of the endpoints of the edge or at the orthogonal projection of $v$ to the supporting line of the edge. 

\subsubsection{Additional predicates}
\begin{figure} \centering\includegraphics[width=0.6\textwidth]{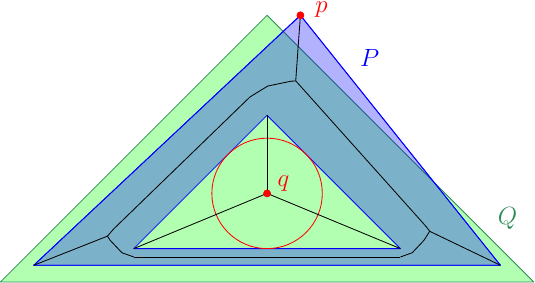}
    \caption{Illustration  of the two cases: The point $p$ on the boundary of $P$ maximizes $\rHausd{p,Q}$. The point $q$ in the interior of $Q$ that is a Voronoi vertex of the edges of $P$ that maximizes $\rHausd{q,P}$.  Figure is best viewed in color.}
    \label{fig:polygon:dist:cases}
\end{figure}
Let $P$ and $Q$ be two polygonal regions that may contain holes. Let further $r\in\RR_+$.
In this section, we give predicates that determine the Hausdorff distance query  $\Hausd{P,Q}\leq r$. The query depends on the two queries for the directed Hausdorff distances $\rHausd{P,Q}\leq r$ and $\rHausd{Q,P}\leq r$. We show how to determine $\rHausd{P,Q}\leq r$. The other direction is analogous. The distance $\rHausd{p,Q}$ for points $p\in P$ can be maximized at points in the interior of $P$ or at points at the boundary of $P$ (see Figure~\ref{fig:polygon:dist:cases} for the two cases). Since the analyses of these two cases are different, we split the query into two general predicates. 

\begin{itemize}
\item $({\mathcal B})$  (Boundary): This predicate returns true if and only if $\rHausd{\partial P,Q}\leq r$.
\item $({\mathcal I})$  (Interior): If $\rHausd{P,Q}\leq r$, this predicate returns true.  In the case that $\rHausd{P,Q} > r$, if $\rHausd{P,Q}>\rHausd{\partial P,Q}$ and $\rHausd{P,Q}> r$, then this predicate returns false. 
\end{itemize}

Note that it is not defined what $({\mathcal I})$ returns if $\rHausd{P,Q}=\rHausd{\partial P,Q}$ and $\rHausd{P,Q}> r$. This does not matter, since the correctness of $\rHausd{P,Q}\leq r$ is still equivalent to both $({\mathcal B})$ and  $({\mathcal I})$ being true.

\begin{figure}[ht] \centering\includegraphics[width=1\textwidth]{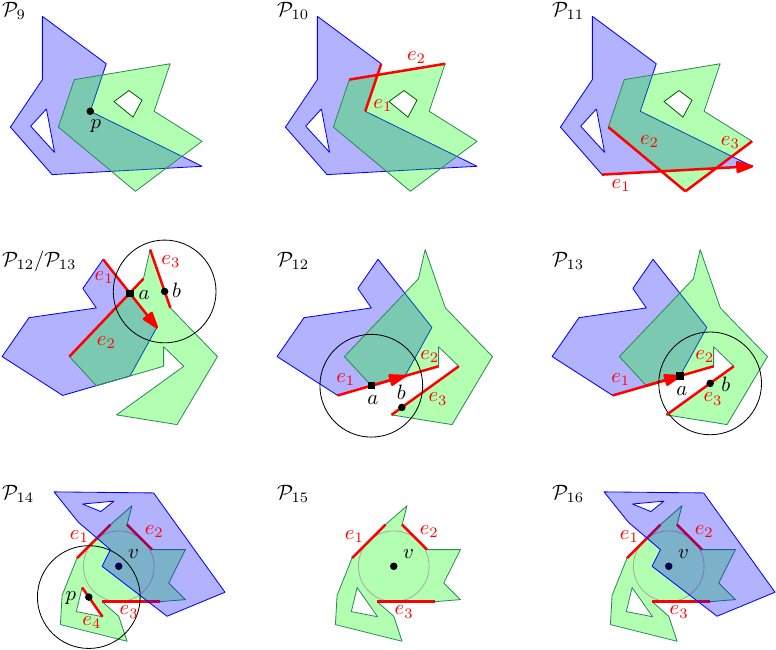}
    \caption{Illustration of the predicates $\Pred_{\ref{hprb1}},\dots, \Pred_{\ref{hpri3}}$ :  In all depicted cases the corresponding predicates are true. Figure is best viewed in color. }
    \label{fig:polygon:predicates}
\end{figure}

Since $({\mathcal B})$ and $({\mathcal I})$ are very general, we define more detailed predicates that can be used to determine feasible truth values of $({\mathcal B})$ and $({\mathcal I})$.  
To determine $({\mathcal B})$, we need the following predicates in combination with $\Pred_{\ref{hpc2}}$ and $\Pred_{\ref{hpc4}}$ defined in Section~\ref{sec:curves:Pred}:
\begin{itemize}
\item \begin{predicate}
Given a vertex $p$ of $P$, this predicate returns true if and only if $p\in Q$.\label{hprb1}
\end{predicate} 
\item \begin{predicate}
Given an edge $e_1$ of $P$ and an edge $e_2$ of $Q$, this predicate is equal to $e_1\cap e_2\neq \emptyset$.\label{hprb2}
\end{predicate} 
\item \begin{predicate}
Given a directed edge $e_1$ of $P$ and two edges $e_2$ and $e_3$ of $Q$, this predicate is true if and only if $e_1\cap e_2\neq \emptyset$, $e_1\cap e_3\neq \emptyset$ and $e_1$ intersects $e_2$ before or at the same point that it intersects $e_3$.\label{hprb3}
\end{predicate} 
\item \begin{predicate}
Given a directed edge $e_1$ of $P$ and two edges $e_2$ and $e_3$ of $Q$, this predicate is true if and only if $e_1\cap e_2\neq \emptyset$ and there exists a point $b$ on $e_3$ such that  $\norm{a-b}\leq r$, where $a$ is the first intersection point in $e_1\cap e_2$.\label{hprb4}
\end{predicate} 
\item \begin{predicate}
Given a directed edge $e_1$ of $P$ and two edges $e_2$ and $e_3$ of $Q$, this predicate is true if and only if $e_1\cap e_2\neq \emptyset$ and there exists a point $b$ on $e_3$ such that  $\norm{a-b}\leq r$, where $a$ is the last intersection point in $e_1\cap e_2$.\label{hprb5}
\end{predicate}   
\end{itemize}
Using Voronoi-vertex-candidates, we  define the detailed predicates for determining $({\mathcal I})$:
\begin{itemize}

\item \begin{predicate}
Given $4$ edges $e_1,e_2,e_3,e_4$ of $Q$ and a point $v$ from the set of Voronoi-vertex-candidates $V(e_1,e_2,e_3)$, this predicate returns true if and only if there exists a point $p\in e_4$, such that $\|v-p\|\leq r.\label{hpri1}$
\end{predicate} 
\item \begin{predicate}
Given $3$ edges $e_1,e_2,e_3$ of $Q$ and a point $v$ from the set of Voronoi-vertex-candidates $V(e_1,e_2,e_3)$, this predicate returns true if and only if $v\in Q$.\label{hpri2}
\end{predicate} 
\item \begin{predicate}Given $3$ edges $e_1,e_2,e_3$ of $Q$ and a point $v$ from the set of Voronoi-vertex-candidates $V(e_1,e_2,e_3)$, this predicate returns true if and only if $v\in P$.\label{hpri3}
\end{predicate}
\end{itemize}
Examples for the predicates $\Pred_{\ref{hprb1}},\dots \Pred_{\ref{hpri3}}$ are depicted in Figure~\ref{fig:polygon:predicates}.

\vspace{9pt}

\begin{restatable}{lemma}{polpreda}\label{lem:hausd:pol:pred1}
For any two polygonal regions $P$ and $Q$ that may contain holes, given the truth values of all predicates of the types $\Pred_{\ref{hpc2}}$, $\Pred_{\ref{hpc4}}$, $\Pred_{\ref{hprb1}}$, $\Pred_{\ref{hprb2}}$, $\Pred_{\ref{hprb3}}$, $\Pred_{\ref{hprb4}}$, and $\Pred_{\ref{hprb5}}$, one can determine the truth value of a predicate of type $({\mathcal B})$.
\end{restatable}

\begin{proof}

    To determine $({\mathcal B})$ it suffices to check for each edge $e$ of $P$ if $\rHausd{e,Q}\leq r$. If this is true for all edges, we return true, otherwise false. Let $e=\overline{uv}$ be an edge of $P$. We first determine which points of $e$ lie outside of $Q$. The $\Pred_{\ref{hprb1}}$ for the vertex $u$, tells us if $u$ lies in $Q$. Checking $\Pred_{\ref{hprb2}}$ and $\Pred_{\ref{hprb3}}$ for $e$ and all edges (respectively pairs of edges) of $Q$ then determines which edges of $Q$ get intersected and in which order they get intersected. Each intersection changes the state of the edge $e$ between lying inside and outside of $Q$. So in total, we get a sequence of intersections of edges with $Q $ where we know for each part between two consecutive intersections, if this part is inside or outside of $Q$. 

    Consider two edges $e_1$ and $e_2$ of $Q$ that intersect $e$ consecutively. Let $s=\overline{s_1 s_2}$ be a subset of $e$ defined by $e_1$ and $e_2$. If $s$ lies inside of $Q$ then we have  $\rHausd{s,Q}= 0$. If $s$ lies outside of $Q$ then we claim that $\rHausd{s,Q}\leq r$ if and only if there exists a sequence of edges $\overline{q_{j_1}q_{j_1+1}}, \overline{q_{j_2}q_{j_2+1}}, \dots, \overline{q_{j_t}q_{j_t+1}}$ for some integer value $t$, such that $\Pred_{\ref{hprb5}}(e,e_1,\overline{q_{j_1}q_{j_1+1}})$ and $\Pred_{\ref{hprb4}}(e, e_2, \overline{q_{j_t}q_{j_t+1}})$ evaluate to true and the conjugate 
    \[ \bigwedge\limits_{i=1}^{t-1} \Pred_{\ref{hpc4}}(e,\overline{q_{j_i}q_{j_i+1}},\overline{q_{j_{i+1}}q_{j_{i+1}+1}})  \]
    evaluates to true. The proof of this claim is analogous to the key part of the proof of Lemma~7.1 in \cite{driemel2021vc}. We include a full proof of the claim here for the sake of completeness.

    Assume such a sequence $\overline{q_{j_1}q_{j_1+1}}, \dots, \overline{q_{j_t}q_{j_t+1}}$ exists. We claim that there exists a sequence of points $a_1, \dots, a_t$ on the line supporting $s$ such that $a_1=s_1$, $a_t=s_2$, and for $1\leq i < t$, $a_i,a_{i+1}\in D_{r}(\overline{q_{j_i}q_{j_i+1}})$. That is, two consecutive points of the sequence are contained in the same stadium. Indeed, for $i=1$, we have that there exists points $a_1,a_2$ with the needed properties because the predicates $\Pred_{\ref{hprb5}}(e,e_1,\overline{q_{j_1}q_{j_1+1}})$ and $\Pred_{\ref{hpc4}}(e,\overline{q_{j_1}q_{j_1+1}},\overline{q_{j_{2}}q_{j_{2}+1}})$ evaluate to true. Likewise, for $i=t-1$, it is implied by the corresponding $\Pred_{\ref{hprb4}}$ and $\Pred_{\ref{hpc4}}$ predicates.  For the remaining $1<i<t-1$, it follows from the corresponding $\Pred_{\ref{hpc4}}$ predicates. This proves our claim.
    
    Now, since each stadium $D_r(\overline{q_{j_i}q_{j_i+1}})$ is
    a convex set, each line segment connecting two consecutive points of
    this sequence $a_i, a_{i+1}$ is contained in one of the stadiums. Note that the set of line
    segments obtained this way forms a connected polygonal curve which fully covers the line segment $s$. It follows that
    \[s\subseteq \bigcup_{0\leq i< t}\overline{a_i a_{i+1}}\subseteq \bigcup_{0\leq i< t} D_{r}(\overline{q_{j_i}q_{j_i+1}}) \]
    Therefore, every point on $s$ is within distance $r$ of some point on $Q$ and thus $\rHausd{s,Q}\leq r$.
    
    For the other direction of the proof, assume that $\rHausd{s,Q}\leq r$. Let $E(Q)$ be the set of all edges of $Q$. The definition of the directed Hausdorff distance implies that 
    \[
    s\subseteq \bigcup_{e\in E(Q)} D_{r}(e) 
    \]
    because $s$ is outside $Q$ and so every point on $s$ must be within a distance $r$ of some point in the boundary of $Q$. Consider the intersections of $s$ with the boundaries of stadiums
    \[s\cap \bigcup_{e\in E(Q)} \partial D_{r}(e).\]
    Let $w$ be the number of intersection points.  Let $t=w+2$. Recall that $s = \overline{s_1s_2}$.  Let $a_1=s_1$.  Let $a_t=s_2$.  Let $a_2,a_3,\ldots,a_{t-1}$ the intersection points between $s$ and the boundaries of the stadiums along $s$ from $s_1$ to $s_2$.
   
    We claim that there exists edges $\overline{q_{j_1}q_{j_1+1}}, \overline{q_{j_2}q_{j_2+1}}, \dots, \overline{q_{j_t}q_{j_t+1}}$ such that the predicates $\Pred_{12}(e,e_1,\overline{q_{j_1}q_{j_1+1}})$, $\Pred_{13}(e,e_2,\overline{q_{j_t}q_{j_t+1}})$, and $\Pred_4(e,\overline{q_{j_i}q_{j_i+1}},\overline{q_{j_{i+1}}q_{j_{i+1}+1}})$ for $1 \leq i < t$ evaluate to true.  Observe that $a_i$ is contained in the intersection of two stadiums for $1 < i < t$ because $a_i$ is the intersection between $s$ and a stadium boundary, and with $s$ is covered by the union of stadiums. Moreover, two consecutive points $a_i, a_{i+1}$ are contained in exactly the same subset of stadiums - otherwise there would be another intersection point with boundary of a stadium in between $a_i$ and $a_{i+1}$. This implies that $\Pred_4(e,\overline{q_{j_i}q_{j_i+1}},\overline{q_{j_{i+1}}q_{j_{i+1}+1}})$ evaluates to true for $1 \leq i < t$. 
    The predicates of types $\Pred_{\ref{hprb4}}$ and $\Pred_{\ref{hprb5}}$ follow trivially from the definitions of $s_1$, $s_2$ and the directed Hausdorff distance. This concludes the proof of the other direction.

    It remains to consider the first and last parts of $e$. Let $s$ be a subset of $e$ defined by its boundaries $s_1,s_2$ where one of the boundaries is one of the vertices $u$ and $v$ of $e$ and the other boundary is the closest intersection of $e$ with an edge of $Q$ or (if this does not exist) the other vertex of $e$. The proof for this case is analogous to the previous case. It only needs predicates of type $\Pred_{\ref{hpc2}}$ for $u$ and $v$ instead of the respective predicates of type $\Pred_{\ref{hprb5}}$ and $\Pred_{\ref{hprb4}}$.
    
\end{proof}

\begin{restatable}{lemma}{polpredb}\label{lem:hausd:pol:pred2}
For any two polygonal regions $P$ and $Q$, given the truth values of all predicates of the types $\Pred_{\ref{hpri1}}$, $\Pred_{\ref{hpri2}}$, and $\Pred_{\ref{hpri3}}$, one can determine the truth value of a predicate of type $({\mathcal I})$.
\end{restatable}

\begin{figure}\centering\includegraphics[width=1\textwidth]{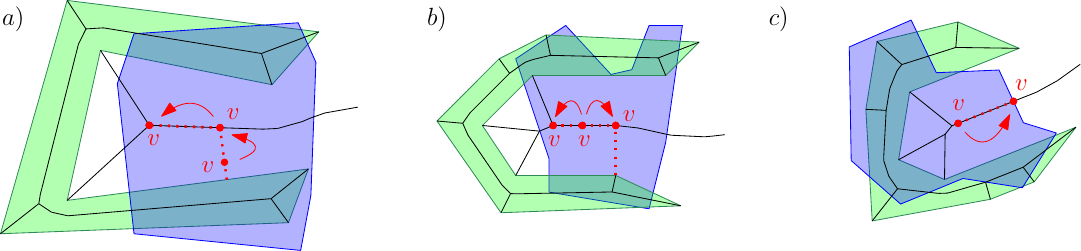}
    \caption{Illustration of the different cases in the proof of Lemma~\ref{lem:hausd:pol:pred2}. It is demonstrated how $v$ can be moved to either increase the distance to $Q$ (a) or to stay in the same distance to $Q$ (b,c). Figure is best viewed in color.}
    \label{fig:voronoi}
\end{figure}

\begin{proof}
We claim that, if $\rHausd{P,Q}>\rHausd{\partial P,Q}$ and $\rHausd{P,Q}>r$, then there has to be a point $v$ in the interior of $P$ that maximizes the Hausdorff distance to $Q$ (i.e. $\rHausd{v,Q}=\max_{p\in P}(\rHausd{p,Q})$) and that is a relevant Voronoi vertex of the edges of $Q$. Before we prove the claim, we show that it implies the statement of the lemma. It follows from the claim that in case of $\rHausd{P,Q}>\rHausd{\partial P,Q}$ and $\rHausd{P,Q}> r$ there is a relevant Voronoi vertex $v$ that lies in $P$ and outside of $Q$ with $\rHausd{v,e}> r$ for all edges $e$ of $Q$. The predicates $\Pred_{\ref{hpri1}}, \Pred_{\ref{hpri2}}$ and $\Pred_{\ref{hpri3}}$ check exactly these properties for a superset of the relevant Voronoi vertices of the edges of $Q$. So, we set $({\mathcal I})$ to false if and only if there is a vertex in this superset that fulfills $\Pred_{\ref{hpri3}}$, does not fulfill $\Pred_{\ref{hpri2}}$, and does not fulfill $\Pred_{\ref{hpri1}}$ for any edge of $Q$. Since all relevant Voronoi vertices are checked, $({\mathcal I})$ will be set to false in all relevant cases. 
On the other hand, if we have $\rHausd{P,Q}\leq r$, then there will be no point that is in $P$, outside of $Q$ and has distance greater $r$ to all edges of $Q$ and $({\mathcal I})$ is set to true.  It remains to show the claim.
    
    We prove the claim by contradiction. We assume that  $\rHausd{P,Q}>r$ and $\rHausd{P,Q}>\rHausd{\partial P,Q}$ and that none of the points in the interior of $P$ that maximize the Hausdorff distance to $Q$ is a relevant Voronoi vertex of the edges of $Q$. Let $v\in P$ be a point maximizing $\rHausd{v,Q}$. Assume that $v$ lies in the Voronoi region of an edge $e$ of $Q$. Then $\rHausd{v,Q}$ can be increased by moving $v$ in perpendicular direction away from $e$ (see Fig.~\ref{fig:voronoi}a)). This would contradict that $v$ maximizies $\rHausd{v,Q}$. So instead, assume that $v$ lies on the Voronoi edge defined by the Voronoi regions of two edges $e_1$ and $e_2$ of $Q$ and that $v$ is not a relevant Voronoi vertex. If $e_1$ and $e_2$ are not parallel, then it can be shown with a straightforward case analysis that there is a direction in which $v$ can be moved  along the Voronoi diagram to increase $\rHausd{v,Q}$ (see Fig.~\ref{fig:voronoi}a)). The direction depends on the the closest points $v_1,v_2$ to $v$ on $e_1,e_2$. If $v_1$ and $v_2$ are perpendicular projections of $v$ to $e_1$ and $e_2$, then $v$ can be moved along the angle bisector of $e_1$ and $e_2$ away from the intersection of $\ell(e_1)$ and $\ell(e_2)$. If only one of $v_1$ and $v_2$ is not a perpendicular projection, then $v$ can be moved along the parabola defined by $v_1$ and $e_2$ (or $v_2$ and $e_1$) in both directions. If both $v_1$ and $v_2$ are not a perpendicular projection, then $v$ can be moved in any direction $d$ with $\langle v-v_2, d \rangle\geq 0$ and $\langle v-v_1, d \rangle\geq 0$. If $e_1$ and $e_2$ are parallel, it can happen that locally $\rHausd{p,Q}$ stays constant for moving $v$ along the Voronoi edge in both directions until either a relevant Voronoi vertex is reached (see Fig.~\ref{fig:voronoi}b)), the boundary of $P$ is reached (see Fig.~\ref{fig:voronoi}c)) or the orthogonal projection of $v$ to either one of the edges $e_1$ and $e_2$ reaches an endpoint of the respective edge (see Fig.~\ref{fig:voronoi}b)). In all three cases, we get a contradiction. The first case contradicts the assumption that there is no relevant Voronoi vertex $v$ that maximizes $\rHausd{v,Q}$, the second case contradicts the assumption that $\rHausd{P,Q}>\rHausd{\partial P,Q}$  and in the third case $\rHausd{p,Q}$ would start increasing and contradict that $v$ maximizes $\rHausd{v,Q}$.
\end{proof}

\section{The predicates are simple}

It remains to show that the predicates defined in Sections~\ref{sec:curves:Pred} and \ref{sec:Hausd:Pred:Pol} can be determined by a polynomial number of simple predicates. Corollary~\ref{cor:vc} then implies that all considered range spaces have $VCdim(\RSpace_{\rho,k})$ in $O(dk\log(km))$.

\subsection{Technical lemmas}\label{sec:genapp}
In this section, we establish some general lemmas that will help us to show that predicates can be determined by a polynomial number of simple predicates. We first introduce some new notation.

\begin{definition}
    Let $d\in\NN$. We call a function $f:\RR^d \rightarrow \RR$ \emph{well behaved} if $f$ is a linear combination of constantly many rational functions of constant degree. Let $\RR^{d_1} \times \cdots \times \RR^{d_i}$ be any partition of $\RR^d$ into cross products, i.e., $\sum_{j=1}^i d_j = d$. For any $X = (x_1,\ldots,x_i) \in \RR^{d_1} \times \cdots \times \RR^{d_i}$, we may use $f(x_1,\ldots,x_i)$ or $f(X)$ to denote $f(x)$, where $x$ is the concatenation $x_1\cdots x_i$.
\end{definition}

For many of our predicates, we have to determine the order of two points on a line. For example, when we check if the intersections of a line with other geometric objects overlap. The following lemma shows that determining the order is simple.

\begin{restatable}{lemma}{obsorder}\label{obs:order}
    Let $d\in\NN$. Let $p$ and $q$ be two points in a finite point set $P\subset\RR^d$.  Consider two points $v$ and $w$ on the line $\ell(\overline{pq})$  given by
    \begin{align*}
        v&= p+t_1(P)(q-p)\\
        w&= p+t_2(P)(q-p)
    \end{align*}
    with  $t_i(P)=a_i(P)+b_i(P)\sqrt{c_i(P)}$
    where $a_i, b_i$ and $c_i$ are well behaved functions for $i\in\{1,2\}$. It is a simple predicate to determine the order of $v$ and $w$ in direction $(q-p)$.
\end{restatable}

Note that the order in Lemma~\ref{obs:order} can be decided by solving $t_1(P)\geq t_2(P)$. So Lemma~\ref{obs:order} directly follows from the following more general lemma.

\begin{restatable}{lemma}{lemsqrt}\label{lem:sqrt}
    Consider the following inequalities:
    \begin{align}
        a(x) &\geq 0, \label{ineq:1}\\
        b\left(x,\sqrt{c(x)},\sqrt{d(x)}\right) &\geq 0, \label{ineq:2}\\
        e\left(x,\sqrt{f(x)},\sqrt{g\left(x,\sqrt{f(x)}\right)}\right)&\geq 0, \label{ineq:3}
    \end{align} 
where $a$, $b$, $c$, $d$, $e$, $f$, and $g$ are well behaved functions. The following statements hold:
    \begin{enumerate}[{\em (i)}]
        \item Inequality (\ref{ineq:1}) is simple. \label{st:1}
        \item Inequality (\ref{ineq:2}) is simple if $c(x)\geq 0$ and $d(x)\geq 0$.\label{st:2}
        \item Inequality (\ref{ineq:3}) is simple if $f(x)\geq 0$ and $g(x,\sqrt{f(x)})\geq 0$.\label{st:3}
    \end{enumerate}
\end{restatable}

Observe that the inequalities $c(x)\geq 0$, $d(x)\geq 0$ and $f(x)\geq 0$ are simple by the first statement and $g(x,\sqrt{f(x)})\geq 0$ is simple by the second statement. 

\begin{proof}[Proof of Lemma~\ref{lem:sqrt}]
    
Consider~(\ref{st:1}). If we multiply both sides of the inequality $a(x)\geq 0$ by the square of the product of all denominators of the rational functions in $a$, then we get an equivalent inequality that only consists of a polynomial of constant degree on the left side and $0$ on the right side. This inequality is by definition simple.

Consider~(\ref{st:2}). If we multiply both sides of the inequality $b\left(x,\sqrt{c(x)},\sqrt{d(x)}\right)\geq 0$ by the square of the product of all denominators of the rational functions in $b$, then we get an equivalent inequality
    \[b_0\left(x,\sqrt{c(x)},\sqrt{d(x)}\right)\geq 0,\]
    where $b_0$ is a polynomial of constant degree. If we rearrange the terms in $b_0$ we get an equivalent inequality
    \begin{equation}
        b_1(x)+b_2(x)\sqrt{c(x)}\sqrt{d(x)}\leq b_3(x)\sqrt{c(x)}+b_4(x)\sqrt{d(x)}, \label{ineq:4}
    \end{equation}
    where $b_1, b_2, b_3$ and $b_4$ are polynomials of constant degree. To show that (\ref{ineq:4}) is simple, we first show that the sign values of both sides of $(4)$ are determined by simple inequalities. To check the sign of the left side  
    \begin{equation}
        b_1(x)+b_2(x)\sqrt{c(x)}\sqrt{d(x)}\geq 0, \label{ineq:5}
    \end{equation}
    we have to check the signs of $b_1(x)$ and $b_2(x)$. Since $b_1$ and $b_2$ are polynomials of constant degree, their signs are determined by a simple inequality. If $\mathrm{sgn}(b_1(x))= \mathrm{sgn}(b_2(x))$ then the truth value of (\ref{ineq:5}) is directly implied. 
    Otherwise, if $b_2(x) \geq 0$ and $b_1(x) < 0$, then by squaring we get
    \begin{equation*}
        b_2(x)^2c(x)d(x)\geq b_1(x)^2.
    \end{equation*}
    Otherwise, if $b_2(x) < 0$ and $b_1(x) \geq 0$, then by squaring we get
    \begin{equation*}
        b_2(x)^2c(x)d(x)\leq b_1(x)^2.
    \end{equation*}
    In either case, we obtain a simple inequality because it has the same form as (\ref{ineq:1}). The check for the sign value of the right side of (\ref{ineq:4}) is analogous. If the sign values of the two sides differ, we get an immediate solution for the truth value of (\ref{ineq:4}). Otherwise, 
    we get one of the following two inequalities:
\begin{equation}
\begin{array}{rcccl}
b_1(x)^2+b_2(x)^2c(x)d(x)+
& & \leq \, \text{or} \, \geq & & b_3(x)^2c(x)+b_4(x)^2d(x)+ \\
2b_1(x)b_2(x)\sqrt{c(x)}\sqrt{d(x)} & &  & &
2b_3(x)b_4(x)\sqrt{c(x)}\sqrt{d(x)}.
\end{array}
\label{ineq:6}
\end{equation}
Multiplying both sides of (\ref{ineq:6}) by the square of the product of all denominators of the rational functions in $c$ and $d$ and then rearranging the terms gives an equivalent inequality
    \begin{equation*}
        b_5(x)+b_6(x)\sqrt{c(x)}\sqrt{d(x)}\geq 0.
    \end{equation*}
    where $b_5$ and $b_6$ are polynomials of constant degree. This inequality is simple as it has the same form as (\ref{ineq:5}). In total, inequality (\ref{ineq:2}) is simple as a constant combination of simple predicates.

Consider (\ref{st:3}). The structure of the proof of the third statement is very similar to the structure of the proof of the second statement. We still include the details here for completeness. 

    If we multiply both sides of inequality $e\left(x,\sqrt{f(x)},\sqrt{g\left(x,\sqrt{f(x)}\right)}\right)\geq 0$ by the square of the product of all denominators of the rational functions in $e$ and rearrange some terms, then we get an equivalent inequality
    \begin{equation}
        e_1(x)+e_2(x)\sqrt{f(x)}\leq e_3(x)\sqrt{g\left(x,\sqrt{f(x)}\right)}+ e_4(x)\sqrt{f(x)}\sqrt{g\left(x,\sqrt{f(x)}\right)} \label{in:3.1}
    \end{equation}
    where $e_1, e_2, e_3$ and $e_4$ are polynomials of constant degree. We again show that the sign values of both sides of (\ref{in:3.1}) are determined by a simple inequality. The check of the left side is analogous to checking inequality (\ref{ineq:5}). To check the right side 
    \begin{equation}
        e_3(x)\sqrt{g\left(x,\sqrt{f(x)}\right)}+ e_4(x)\sqrt{f(x)}\sqrt{g\left(x,\sqrt{f(x)}\right)}\geq 0 \label{in:3.2}
    \end{equation}
    we have to check the signs of $e_3(x)$ and $e_4(x)$. Since $e_3$ and $e_4$ are polynomials of constant degree, their signs are determined by a simple inequality. If $\mathrm{sgn}(e_3(x))= \mathrm{sgn}(e_4(x))$ then the truth value of (\ref{ineq:5}) is directly implied. Otherwise, we can square both sides of 
    \[-e_3(x)\sqrt{g\left(x,\sqrt{f(x)}\right)}\leq e_4(x)\sqrt{f(x)}\sqrt{g\left(x,\sqrt{f(x)}\right)}\] to get that (\ref{in:3.2}) is equivalent to
    \begin{equation}
        e_3(x)^2g\left(x,\sqrt{f(x)}\right)\leq e_4(x)^2f(x)g\left(x,\sqrt{f(x)}\right).
    \end{equation}
    After rearranging, this is a simple inequality because it has the same form as (\ref{ineq:2}). If the sign values of the two sides of inequality (\ref{ineq:3})  differ, we get an immediate solution for it. Otherwise, we get the following equivalent inequality by squaring both sides:
\begin{align}
&~~~~~e_1(x)^2+e_2(x)^2f(x)+2e_1(x)e_2(x)\sqrt{f(x)} \nonumber \\
& \leq~e_3(x)^2g\left(x,\sqrt{f(x)}\right)+ e_4(x)^2g\left(x,\sqrt{f(x)}\right)+ \label{in:3.3}\\
&~~~~~e_3(x)e_4(x)\sqrt{f(x)}g\left(x,\sqrt{f(x)}\right). \nonumber
\end{align}
    Rearranging the terms in (\ref{in:3.3}) gives an inequality
    \[e_0(x,\sqrt{f(x)})\geq 0\]
    where $e_0$ is a well behaved function. Since this inequality is a special case of inequality (\ref{ineq:2}) it is simple.
\end{proof}

In general, we have to deal with different types of points as part of the predicates. We classify them in the following way.

\vspace{9pt}

\begin{definition}\label{def:root}
    Let $d\in\NN$.  Let $r$ be a positive value. Let $p$ and $q$ be two points in a finite point set $P \subset \RR^d$.  Let $v$ be any point in $\RR^d$.  We say that $v$ is a point of \emph{root-type} \emph{1}, \emph{2} or \emph{3} with respect to $P$ if and only if the coordinates of $v=(v_1,\dots,v_d)$ can be written respectively as
\begin{enumerate}[{\em (i)}]
    \item $v_i = a_i(P)$,
    \item $v_i= b_i\left(P,\sqrt{c(P)},\sqrt{d(P)}\right)$, or
    \item $v_i= e_i\left(P,\sqrt{f(P)},\sqrt{g\left(P,\sqrt{f(P)}\right)}\right)$.   
\end{enumerate}
such that $c$, $d$, $f$, $g$, and $a_i,b_i,e_i$ for $1 \leq i \leq 3$ are well behaved functions with $c(P)\geq 0$, $d(P)\geq 0$, $f(P)\geq 0$, and $g\left(P,\sqrt{f(P)}\right)\geq 0$. 
\end{definition}

\vspace{9pt}

\begin{restatable}{lemma}{lemvertexcandidates}\label{lem:vertex_candidates}

Let $a=\overline{a_1 a_2}, b=\overline{b_1 b_2}$ and $c=\overline{c_1 c_2}$ be edges of a polygonal region that may contain holes. Let $P=\{a_1,a_2,b_1,b_2,c_1,c_2\}$. All points in the set of Voronoi-vertex-candidates $V(a,b,c)$ are of root-type 1, 2 or 3 with respect to $P$. There is only a constant number of combinations of well behaved functions that can define a point in $V(a,b,c)$ (by using these functions as the well behaved functions in Definition~\ref{def:root}). These combinations are uniquely determined by $a,b$ and $c$. For each combination, it is a simple predicate to check if it defines a point in $V(a,b,c)$.
\end{restatable}

\begin{proof}
     Consider the sets $A=\{a_1,a_2,\ell(a)\}$, $B=\{b_1,b_2,\ell(b)\}$ and $C=\{c_1,c_2,\ell(c)\}$. Let $X\in A$, $Y\in B$ and $Z\in C$. This combination of $X,Y$ and $Z$ only contributes points to $V(a,b,c)$ if neither of $X,Y$ and $Z$ is a subset of one of the others. This can be checked with a simple predicate by Lemma~\ref{lem:sqrt}: To check if two points $(u_1,u_2)$ and $(v_1,v_2)$ are the same, we need to check if $u_1=v_1$ and $u_2=v_2$. Checking an equation $(=)$ can be done by checking both inequalities $(\leq)$ and $(\geq)$. Checking if a point $(u_1,u_2)$ lies on a line $\ell(\overline{vw})$ with  $v=(v_1,v_2)$ and $w=(w_1,w_2)$ can be done by checking if there exists a $t$ such that
     \[\begin{pmatrix} v_1 \\ v_2 \end{pmatrix} + t \begin{pmatrix} w_1-v_1 \\ w_2-v_2 \end{pmatrix} = \begin{pmatrix} u_1 \\ u_2 \end{pmatrix}.\]
     This is equivalent to the following equation being true.
     \[v_2+\frac{u_1-v_1}{w_1-v_1}(w_2-v_2)=u_2\]
     To check if two lines $\ell(\overline{pq})$ and $\ell(\overline{vw})$ coincide, we have to check if $p$ is on the line $\ell(\overline{vw})$ as before and additionally if the lines have the same slope. This can be done by checking the truth value of $\frac{p_2-q_2}{p_1-q_1}=\frac{v_2-w_2}{v_1-w_1}$.
     
     If the checks determine that one of $X,Y$ and $Z$ is the subset of one of the others, then all combinations of well behaved functions based on the combination $X,Y,Z$ can be ignored. Otherwise, we have 4 different cases for the types of objects $X,Y,Z$. It can be that there are 3 points, 2 points and 1 line, 1 point and 2 lines or 3 lines. Consider the biscetors of the pairs $(X,Y)$, $(X,Z)$ and $(Y,Z)$. The Voronoi-vertex-candidates are the intersections of these bisectors. It suffices to find  the intersections of two bisectors, since the third one intersects the other two in the same points (by definition).

     \textbf{Case 1 (3 points):} The bisector of the 3 points $u=(u_1,u_2)$, $v=(v_1,v_2)$ and $w=(w_1,w_2)$ intersect if the points are not colinear. This is just a simple predicate to check, as we have seen before by checking if $u$ lies on the line $\ell(\overline{vw})$. Assume $u$, $v$ and $w$ are not colinear. Then the bisector of $u$ and $v$ parameterized in $s$ is given by
     \[\ell_1(s)=\frac{1}{2}\begin{pmatrix} u_1+v_1 \\ u_2+v_2 \end{pmatrix} + s \begin{pmatrix} v_2-u_2 \\ u_1-v_1 \end{pmatrix}\]
     and the bisector of $w$ and $v$ parameterized in $t$ is given by
     \[\ell_2(s)=\frac{1}{2}\begin{pmatrix} w_1+v_1 \\ w_2+v_2 \end{pmatrix} + t \begin{pmatrix} v_2-w_2 \\ w_1-v_1 \end{pmatrix}.\]
     So if we set $\ell_1(s)=\ell_2(t)$, we get two linear equations with the two variable $s$ and $t$ of the form
     \[f(P)+g(P)s+h(P)t=0\]
     where $f$, $g$, and $h$ are well behaved functions. Since the points were not colinear, the solution for $t$ is a uniquely determined well behaved function and therefore the intersection point of the bisectors is a point of root-type 1.

     \textbf{Case 2 (2 points, 1 line):} Note that a line between two points in $P$ can be written as
     \[f_1(P)x+f_2(P)y+f_3(P)=0\]
     where $f_1$, $f_2$, and $f_3$ are well behaved functions. The bisector between such a line and a point $(u_1,u_2)$ is given by the parabola
     \begin{equation}
         \frac{f_1(P)x+f_2(P)y+f_3(P)}{f_1(P)^2+f_2(P)^2}=(x-u_1)^2+(y-u_2)^2.\label{eq:parabola}
     \end{equation}
     For the bisector of two points $u=(u_1,u_2)$ and $(v_1,v_2)$ parameterized in $t$, we have as before
     \begin{align}
         x &= \frac{1}{2}(u_1+v_1)+t(v_2-u_2) \label{eq:bisec:x}\\
         y &= \frac{1}{2}(u_2+v_2)+t(u_1-v_1) \label{eq:bisec:y}
     \end{align}
     Inserting (\ref{eq:bisec:x}) and (\ref{eq:bisec:y}) into (\ref{eq:parabola}) gives a quadratic equation in $t$ with solutions of the form
     \[t_{1/2}=g_1(P)\pm\sqrt{g_2(P)}\]
     where $g_1$ and $g_2$ are well behaved functions. If $g_2(P)<0$ then there is no intersection. This can be checked with a simple predicate by Lemma~\ref{lem:sqrt}. Otherwise, the (up to two) intersections of the two bisectors are points of root-type 2.

     \textbf{Case 3 (1 point, 2 lines):}  In this case it can happen that the 2 lines are parallel. We have already shown that it can be checked by a simple predicate if the two lines have the same slope. Let the two lines be $\ell(\overline{pq})$ and $\ell(\overline{vw})$ with $p=(p_1,p_2)$, $q=(q_1,q_2)$, $v=(v_1,v_2)$ and $w=(w_1,w_2)$.

     \textbf{Case 3.1 (2 lines are parallel):} The bisector of $\ell(\overline{pq})$ and $\ell(\overline{vw})$ parameterized in $t$ is given by
        \begin{align*}
         x &= \frac{1}{2}(p_1+v_1)+t(p_1-q_1) \\
         y &= \frac{1}{2}(p_2+v_2)+t(p_2-q_2) 
     \end{align*}
     So analogous to Case 2, the existence of an intersection of such a bisector with a parabola of the form (\ref{eq:parabola}) is a simple predicate and if an intersection exists, the (up tp two) intersections are points of root-type 2.

     \textbf{Case 3.2 (2 lines are not parallel):} The bisector of $\ell(\overline{pq})$ and $\ell(\overline{vw})$ is the union of their two angle bisectors. The angle bisectors are uniquely determined by the intersection point of $\ell(\overline{pq})$ and $\ell(\overline{vw})$ and their slopes. Analogous to Case~1, it can be seen that the intersection of $\ell(\overline{pq})$ and $\ell(\overline{vw})$ is a point $(g_1(P),g_2(P))$ where $g_1$ and $g_2$ are well behaved functions. Let $m'$ and $m''$ be the slopes of $\ell(\overline{pq})$ and $\ell(\overline{vw})$.  The angle between two lines with slope $m'$ and $m''$ is given by $\tan^{-1}(\frac{m'-m''}{1+m'm''})$.  Since the angle bisectors have the same angle to both of the lines just with different sign, we get for the slope $m$ of an angle bisector that
     \[\frac{m-m'}{1+mm'}=-\frac{m-m''}{1+mm''}\]
     Solving this equation for $m$ gives two solutions of the form
     \[m_{1/2}=g_3(P)\pm \sqrt{g_4(P)}\]
     where $g_3$ and $g_4$ are well behaved functions with $g_4(P)\geq 0$. So in total the angle bisectors are given by
     \begin{align}
         x &= g_1(P)+t \label{eq:angle:x}\\
         y &= g_2(P)+t (g_3(P))\pm \sqrt{g_4(P)} \label{eq:angle:y}
     \end{align}
     For each of the angle bisectors, inserting (\ref{eq:angle:x}) and (\ref{eq:angle:y}) in (\ref{eq:parabola}) gives a quadratic equation in $t$ of the form 
     \[t^2h_1(P,\sqrt{g_4(P)})+th_2(P,\sqrt{g_4(P)})+h_3(P,\sqrt{g_4(P)})\]
     where $h_1$, $h_2$, and $h_3$ are well behaved functions. The solutions for $t$ therefore have the form
     \[t_{1/2}=h_4(P,\sqrt{g_4(P)})\pm \sqrt{h_5(P,\sqrt{g_4(P)})}\]
     where $h_4$ and $h_5$ are well behaved functions. If $h_5(P,\sqrt{g_4(P)})<0$, then there is no intersection. This is simple by Lemma~\ref{lem:sqrt}. Otherwise, the (up to two) intersections are points of root-type 3. In total, there can be up to four intersections because there are two angle bisectors.

     \textbf{Case 4 (3 lines):} As we have seen before, all occurring bisectors are unions of (up to two) lines of the form given in (\ref{eq:angle:x}) and (\ref{eq:angle:y}). Note that the bisector of two parallel lines can also be realized in that way by setting $g_4(P)=0$. Consider the intersection of two of these bisectors $\ell_1(s)$ and $\ell_2(t)$:
     \[\ell_1(s)=\begin{pmatrix} f_1(P) \\ f_2(P) \end{pmatrix} + s \begin{pmatrix} 1 \\ f_3(P)+\sqrt{f_4(P)} \end{pmatrix}\] and
     \[\ell_2(t)=\begin{pmatrix} g_1(P) \\ g_2(P) \end{pmatrix} + t \begin{pmatrix} 1 \\ g_3(P)+\sqrt{g_4(P)} \end{pmatrix},\]
where $f_i$ and $g_i$ for $1 \leq i \leq 4$ are well behaved functions, $f_4(P)\geq0$, and $g_4(P)\geq0$. If we set $\ell_1(s)=\ell_2(t)$, we get a system of two linear equations in $s$ and $t$. The system has a unique solution if $f_3(P)+\sqrt{f_4(P)}\neq g_3(P)+\sqrt{g_4(P)}$.  This inequality is simple by Lemma~\ref{lem:sqrt}. Otherwise, there are two parallel bisectors. These can only intersect if they are the same.  It would imply that two of $X,Y$ and $Z$ are also the same. This case was already taken care of in the beginning of the proof, so we only need to consider the case that $f_3(P)+\sqrt{f_4(P)}\neq g_3(P)+\sqrt{g_4(P)}$. In this case, there is a unique solution for $t$ of the form
     \[t=h(P,\sqrt{f_4(P)},\sqrt{g_4(P)})\]
     where $h$ is a well behaved function. So the intersection is a point of root-type 2. The proof is analogous if one or two of the considered bisectors have a minus in (\ref{eq:angle:y}).

\end{proof}

A reoccurring predicate is the decision if a given point is within a given distance to a given edge. We show in the following lemma that such predicates are simple.

\vspace{9pt}

\begin{restatable}{lemma}{lempointedge}\label{lem:point:edge}
Let $d\in\NN$.  Let $r$ be a positive value. Let $p$ and $q$ be two points in a finite point set $P\subset\RR^d$. 
Let $v$ be an arbitrary point in $\RR^d$. Let $\Pred$ be the predicate to decide if there exists a point $u\in \overline{pq}$ such that $\norm{u-v}\leq r$. If $v$ is a point of root-type 1, 2, or 3 with respect to $P$, then $\Pred$ is simple.
\end{restatable}

\begin{proof}
    The truth value of the predicate $\Pred$ can be determined by checking if $v$ is in the stadium $D_{r}(\overline{pq})$. For this check, it suffices to check if  $v$ is in at least one of $B_{r}(p)$, $B_{r}(q)$ and $R_{r}(\overline{pq})$. For $B_{r}(p)$ and $B_{r}(q)$ ,we have to check the inequalities
    \begin{align}
        r^2 -\norm{v-p}^2&\geq 0 \label{ieq:pe:1}\\
        r^2 -\norm{v-q}^2&\geq 0. \label{ieq:pe:2}
    \end{align}
     To check if $v\in R_{r}(\overline{pq})$, consider the closest point $s$ to $v$ on the line $\ell(\overline{pq})$. The truth value of 
    \begin{equation}
        r^2-\norm{s-v}^2\geq 0 \label{ieq:pe:3}
    \end{equation}
    uniquely determines whether $v$ is in the cylinder $C_{r}(\overline{pq})$. The truth values of 
    \begin{align}
       \norm{p-q}^2-\norm{p-s}^2&\geq 0, \label{ieq:pe:4}\\
       \norm{p-q}^2-\norm{q-s}^2&\geq 0 \label{ieq:pe:5}
    \end{align}
    further determine if $s$ is on the edge $\overline{pq}$. So the truth values of the inequalities (\ref{ieq:pe:3}), (\ref{ieq:pe:4}) and (\ref{ieq:pe:5}) determine the truth value of $v\in R_{r}(\overline{pq})$.
    The closest point to $v$ on the line $\ell(\overline{pq})$ is
    \[s= p+\frac{(p-q)\langle p-q, v-p \rangle}{\norm{p-q}^2}.\]
    For each coordinate of $s$, we have
    \[s_j=p_j+(p_j-q_j)\frac{\sum_{i=1}^d (p_i-q_i)(v_i-p_i)}{\sum_{i=1}^d (p_i-q_i)^2}.\]
    Note that for any two points $x,y\in\RR^d$,
    \[\norm{x-y}^2=\sum_{i=1}^d (x_i-y_i)^2\] is a polynomial of constant degree. So for each of the inequalities (\ref{ieq:pe:1}), (\ref{ieq:pe:2}), (\ref{ieq:pe:3}), (\ref{ieq:pe:4}), and (\ref{ieq:pe:5}), if we insert all coordinates of $v$ into that inequality and rearrange the terms, then we get an equivalent inequality of one of the following types depending on the root-type of $v$:
    \begin{align*}
        h_1(P) &\geq 0,\\
        h_2\left(P,\sqrt{c(P)},\sqrt{d(P)}\right)&\geq 0,\\
        h_3\left(P,\sqrt{f(P)},\sqrt{g\left(P,\sqrt{f(P)}\right)}\right)&\geq 0,
    \end{align*}
    where $c,d,f, g, h_1, h_2$ and $h_3$ are well behaved functions. By Lemma~\ref{lem:sqrt}, all three types of inequalities are simple. So, in all three cases, only a constant number of simple inequalities have to be checked to determine $\Pred$. Therefore $\Pred$ is simple.  
\end{proof}  

Many of our predicates depend on the intersections of geometric objects. We address in the next lemmas that these intersections have nice properties and that the existence of these intersections can be determined by a simple predicate.

\vspace{9pt}

\begin{restatable}{lemma}{lemrayedge}\label{lem:ray:edge}
 Let $p=(p_1,p_2)$ and $q=(q_1,q_2)$ be two points in a finite point set $P\subset\RR^2$.  Let $hr(v)$ be a horizontal ray that shoots from a point $v \in \RR^2$ to the right.  Let $\Pred$ be the predicate to decide if $hr(v)\cap\overline{pq}\neq \emptyset$. If $v$ is a point of root-type 1, 2, or 3 with respect to $P$, then $\Pred$ is simple.  
\end{restatable}

\begin{proof}
To check if $hr(v)$ intersects the line $\ell(\overline{pq})$, one can first check whether $p_2-q_2=0$ by testing the simple inequalities $p_2- q_2\geq 0$ and $p_2- q_2\leq 0$. If this is the case, then an intersection is still possible if $v_2-p_2=0$. The inequalities $v_2- p_2\geq 0$ and $v_2- p_2\leq 0$ are also  simple by Lemma~\ref{lem:sqrt}. The root-type of $v$ determines which case of the lemma to use. If $v_2- p_2 = 0$ is also true, then it can be determined whether $hr(v)\cap\overline{pq}\neq \emptyset$ by checking $v_1-p_1\geq 0$ and $v_1-q_1\geq 0$. These inequalities determine the relative positions of $v$ to $p$ and $q$ on the horizontal line. They are again simple by Lemma~\ref{lem:sqrt}.

If $p_2\neq q_2$, then the intersection of the horizontal line through $v$ and the line $\ell(\overline{pq})$ is a uniquely defined point $s=p+t(p-q)$ with $t=\frac{(v_2-p_2)}{(p_2-q_2)}$. In this case, it remains to check if $1\geq t$ and $t\geq 0$ to see if $s$ lies on the edge $\overline{pq}$ and to check if $s_1\geq v_1$ to see if $s$ lies on the right side of $v$ and is on the ray $hr(v)$. The inequalities $1\geq t$, $t\geq 0$ and $s_1\geq v_1$ are simple by Lemma~\ref{lem:sqrt} (Rearrange and choose the case of the lemma based on the root-type of $v$).
\end{proof}

\vspace{6pt}

\begin{restatable}{lemma}{lemedgeedge}\label{lem:edge:edge}
 Let $p$, $q$, $u$, and $v$ be four points in a finite point set $P\subset\RR^2$.  There is a simple predicate $\Pred$ that decides if $\overline{pq}\cap\overline{uv}\neq \emptyset$. If $\overline{pq} \cap \overline{uv} \not= \emptyset$, it is either a uniquely defined point $s$ given by 
 \[s= p+t(P)(q-p),\]
where $t$ is a well behaved function, or an edge $\overline{xy}$ where $x,y\in \{p,q,u,v\}$. In case that $\overline{pq} \cap \overline{uv}$ is an edge, there is a simple predicate to decide if a given pair of points $x,y\in \{p,q,u,v\}$ define the intersection. 
\end{restatable}

\begin{proof}
    We can write the line $\ell(\overline{pq})$ as $p+t(p-q)$ parameterized in $t$ and the line $\ell(\overline{uv})$ as $u+t'(v-u)$ parameterized in $t'$. The intersection of the lines is therefore defined by the solutions of the system of linear equations
    \begin{align*}
        p+t(p-q)=u+t'(v-u)
    \end{align*}
     which is equivalent to
     \begin{align*}
        t(p-q)+t'(u-v) +(p-u)=0.
    \end{align*}
    The above is a system of two linear equations with two variables $t, t'$ of the form
    \[a_i t +b_i t' + c_i= 0\]
    where $a_i=(q_i-p_i)$, $b_i=(u_i-v_i)$ and $c_i=(p_i-u_i)$ for $i\in\{1,2\}$. This system has a unique solution if $\frac{a_1}{a_2}\neq \frac{b_1}{b_2}$, no solution $\frac{a_1}{a_2}= \frac{b_1}{b_2}\neq \frac{c_1}{c_2}$ and an infinite number of solutions if $\frac{a_1}{a_2}= \frac{b_1}{b_2}= \frac{c_1}{c_2}$. Each of these relations can be checked by replacing $=$ (or $\neq$) with $\leq$ and $\geq$ and checking both inequalities. So the existence of an intersection can be checked by checking a constant number of simple inequalities. 
    
    Note that the coefficients of the linear equations are linear combinations of coordinates of points in $P$. So, if the system has a unique solution, the solution for $t$ can be written as a well behaved function with input $P$. In this case, it still remains to check $t\geq 0$ and $t\leq 1$ to see if the intersection is on the edge $\overline{pq}$. By Lemma~\ref{lem:sqrt}, these are simple inequalities.
    
    If the system does have an infinite number of solutions, the lines $\ell(\overline{pq})$ and $\ell(\overline{uv})$ must coincide. In this case, the  solutions $t_u$ and $t_v$ of the equations $p_1+t_u(q_1-p_1)=u_1$ and $p_1+t_v(q_1-p_1)=v_1$ are uniquely determined values that can be written as well behaved functions with input $P$. Comparing $t_1,t_2,0$ and $1$ decides if the edges $\overline{pq}$ and $\overline{uv}$ intersect and which points $x,y\in\{p,q,u,v\}\subseteq P$ determine the intersection $\overline{xy}$ (if it exists). Since $t_1$ and $t_2$ are well behaved functions with input $P$, each comparison is a simple predicate.
\end{proof}

\vspace{6pt}

\begin{restatable}{lemma}{lemballline}\label{lem:ball:line}
Let $d\in\NN$.  Let $r$ be a positive value. Let $p$, $q$, and $v$ be three points in a finite point set $P\subset\RR^d$. There is a simple predicate $\Pred$ that decides if $\ell(\overline{pq})\cap B_{r}(v)\neq \emptyset$.  If $\ell(\overline{pq}) \cap B_{r}(v) \not= \emptyset$, then the first and the last point of the intersection in direction $(q-p)$ are uniquely defined by
\[
s_i = p+t_i(P)(q-p) \quad \text{for $i \in \{1,2\}$},
\]
where $t_1(P) = f(P) + \sqrt{g(P)}$ and $t_2(P) = f(P) - \sqrt{g(P)}$ for some well behaved functions $f$ and $g$. 
\end{restatable}

\begin{proof}
We can write the line $\ell(\overline{pq})$ as $p+t(p-q)$ parameterized in $t$. The intersection $\ell(\overline{pq}) \cap B_{r}(v)$ is therefore defined by the solutions of 
\begin{equation*}
   \norm{p+t(q-p)-v}^2 \leq r^2 \, \iff \,
   \sum_{i=1}^d (t(q_i-p_i)+(p_i-v_i))^2 \leq r^2.
\end{equation*}
This inequality is equivalent to a quadratic equation of the form $t^2+at+b\leq 0$, where 
\[a=\frac{2\sum_{i=1}^d(p_i-v_i)(q_i-p_i)} {\sum_{i=1}^d(q_i-p_i)^2}\quad \text{and} \quad b=\frac{\sum_{i=1}^d (p_i-v_i)^2 - r^2}{\sum_{i=1}^d(q_i-p_i)^2}.\]
We therefore have $t_1 = -\frac{a}{2}+\sqrt{\frac{a^2}{4}-b}$ and $t_2 = -\frac{a}{2}-\sqrt{\frac{a^2}{4}-b}$  as long as $\frac{a^2}{4}-b\geq 0$.  If we have $\frac{a^2}{4}-b< 0$ then the intersection is empty. By Lemma~\ref{lem:sqrt}(i), this inequality is simple.
\end{proof}

\vspace{6pt}

\begin{restatable}{lemma}{lemcapcylline}\label{lem:capcyl:line}
Let $d\in\NN$.  Let $r$ be a positive value. Let $p$, $q$, $u$, and $v$ be four points in a finite point set $P\subset\RR^d$.  
\begin{itemize}
\item There is a simple predicate $\Pred$ that decides if $\ell(\overline{pq})\cap R_{r}(\overline{uv})\neq \emptyset$.

\item If $\ell(\overline{pq}) \cap R_{r}(\overline{uv}) \not= \emptyset$, then the first and the last points of the intersection in direction $(q-p)$ are given by
\[
s_i = p+t_i(P)(q-p) \quad \text{for $i \in \{1,2\}$},
\]
where $t_i(P)=f_i(P)+h_i(P)\sqrt{g_i(P)}$ for some well behaved functions $f_i$, $g_i$, and $h_i$.  

\item If $\ell(\overline{pq}) \cap R_{r}(\overline{uv}) \not= \emptyset$, there exists a constant number of candidates for the first and the last points of the intersection in direction $(q-p)$.  These candidates are uniquely determined by equations in $p$, $q$, $u$, $v$, and $r$.  There is a simple predicate to decide which of these candidates are the first and last points in the intersection.
\end{itemize}
\end{restatable}

\begin{proof} This proof of Lemma~\ref{lem:capcyl:line} is based on the proof of Lemma~7.2 in \cite{driemel2021vc} that uses similar arguments.  We can write the line $\ell(\overline{pq})$ as $p+t(p-q)$ parameterized in $t$ and the line $\ell(\overline{uv})$ as $u+t'(v-u)$ parameterized in $t'$. We can determine the intersection of $\ell(\overline{pq})$ and $R_{r}(\overline{uv})$ by intersecting $\ell(\overline{pq})$ with the boundary of the infinite cylinder $C_{r}(\overline{uv})$ and the two limiting hyperplanes $P(\overline{uv})$ and $P(\overline{vu})$. 

The intersection of $\ell(\overline{pq})$ with the boundary of $C_{r}(\overline{uv})$ is defined by all pairs $(t,t')$ that fulfill the equality
\begin{align}
   &\quad\quad\quad\norm{p+t(q-p)-u-t'(v-u)} = r \nonumber \\
   &\iff \sum_{i=1}^d ((p_i-u_i)+t(q_i-p_i)+t'(u_i-v_i))^2-r^2 = 0 \label{ineq:cylinder}
\end{align}
For any fixed $t$ the above equation is a quadratic equation in $t'$ where the discriminant is a quadratic equation in $t$ of the form
\[a(P)t^2+b(P)t+c(P)\]
where $a$, $b$ and $c$ are well behaved functions. If the discriminant is equal to $0$, then equation~(\ref{ineq:cylinder}) has exactly one solution. This is only the case for points $p + t(p-q)$ on the boundary of $C_{r}(\overline{uv})$ because the balls of radii $r$ around such points intersect $\ell(\overline{uv})$ exactly once. If $a(P)=b(P)=c(P)=0$, all points of $\ell(\overline{pq})$ are on the boundary of $C_{r}(\overline{uv})$ and the intersection of the boundary of $C_{r}(\overline{uv})$ and $\ell(\overline{pq})$ therefore consists of the whole line $\ell(\overline{pq})$. The truth value of $a(P)=b(P)=c(P)=0$ can be checked by checking $a(P)\geq 0$, $a(P)\leq 0$, $b(P)\geq 0$, $b(P)\geq 0$, $c(P)\leq 0$ and $c(P)\leq 0$ which are simple by Lemma~\ref{lem:sqrt}(i).

If the intersection is finite, the solutions $s_1$ and $s_2$ for $a(P)t^2+b(P)t+c(P)=0$ define the intersection points of the boundary of $C_{r}(\overline{uv})$ and $\ell(\overline{pq})$. We have
\[s_i = -\frac{b(P)}{2a(P)} + (-1)^i \sqrt{\frac{b(P)^2}{4a(P)^2}-\frac{c(P)}{a(P)}} \quad \text{for $i \in \{1,2\}$}, \]
as long as $\frac{b(P)^2}{4a(P)^2}-\frac{c(P)}{a(P)}\geq 0$. If we have $\frac{b(P)^2}{4a(P)^2}-\frac{c(P)}{a(P)}<0$, then the intersection is empty. By Lemma~\ref{lem:sqrt}(i), this inequality is simple.

The intersection of $\ell(\overline{pq})$ with $P(\overline{uv})$ is given by all parameters $z\in\RR$ such that
\[
  \langle p+z(q-p)-u, v-u\rangle = 0 
  \, \iff \, \langle p-u, v-u \rangle + z \langle q-p, v-u \rangle = 0.
\]
It is possible that the whole line is contained in the hyperplane, or the intersection is empty, or the intersection is only one point.  The line $\ell(\overline{pq})$ is parallel to $P(\overline{uv})$ if and only if $\langle p-u, v-u \rangle=0$.  In this case, $\ell(\overline{pq}) \subset P(\overline{uv})$ if and only if $\langle p-u, v-u \rangle=0$. By replacing $=$ with $\leq$ and $\geq$ we can get a constant number of simple inequalities that checks whether $\ell(\overline{pq}) 
\subset P(\overline{uv})$, or $\ell(\overline{pq}) \cap P(\overline{uv}) = \emptyset$, or $\ell(\overline{pq}) \cap P(\overline{uv})$ is a unique point (simple by Lemma~\ref{lem:sqrt}). If the intersection is a unique point, it is given by the parameter
\[z_u=-\frac{\langle p-u, v-u \rangle}{\langle q-p, v-u \rangle}\]
The intersection with $P(\overline{vu})$ is analogous.  In the case that $\ell(\overline{pq}) \cap P(\overline{vu})$ is a unique point, the parameter is
\[z_v=-\frac{\langle p-v, v-u \rangle}{\langle q-p, v-u \rangle}.\]
To check if the parameters $z_u$ and $z_v$ define points on $R_{r}(\overline{uv})$, we check 
\[\norm{z_u - u}^2\leq r^2\quad \text{and} \quad \norm{z_v - v}^2\leq r^2\]
which are simple by Lemma~\ref{lem:point:edge} where we choose $\overline{uu}$ (respectively $\overline{vv}$) as the degenerate edge that just consists of one point. Comparing $s_1,s_2,z_u$ and $z_v$ decides  which points determine the intersection of $\ell(\overline{pq})$ and $C_{r}(\overline{uv})$ (if the intersection exists). Each comparison is a simple predicate by Lemma~\ref{lem:sqrt}(i).
\end{proof}

\subsection{Predicates for polygonal curves}\label{sec:proof:Combined}

In this section we show that the predicates $\Pred_{\ref{hpc1}},\dots,\Pred_{\ref{fpc4}}$ are simple. 

\vspace{9pt}

\begin{lemma}\label{lem:p3478:eff}
For any two polygonal curves $P\in\XX^{d}_m, Q\in\XX^d_k$ and a radius $r\in\RR_+$, if we treat $P$ and $Q$ as points in $\RR^{dm}$ and $\RR^{dk}$, respectively, by concatenating their vertex coordinates, then each of the predicates of types $\Pred_{\ref{hpc1}}$, $\Pred_{\ref{hpc2}}$, $\Pred_{\ref{hpc3}}$, and $\Pred_{\ref{hpc4}}$ is a simple function that maps $(P,Q,r) \in \RR^{\dim m} \times \RR^{dk+1}$ to $\{0,1\}$ 
\end{lemma}
\begin{proof}
For $\Pred_{\ref{hpc1}}$ and $\Pred_{\ref{hpc2}}$, this statement directly follows from Lemma~\ref{lem:point:edge}.
Let $\Pred$ be a predicate of type $\Pred_{\ref{hpc3}}$ or $\Pred_{\ref{hpc4}}$ with input $((P,Q),r)$. $\Pred$ can be determined by checking if a line $\ell(\overline{pq})$ intersects a double stadium  $D_{r,2}(\overline{uv},\overline{xy})$ for some points $p,q,u,v,x,y\in P\cup Q$. For $\Pred=\Pred_{\ref{hpc3}}$, we have $\overline{pq}=\overline{q_iq_{i+1}}$, and for $\Pred=\Pred_{\ref{hpc4}}$, we have $\overline{pq}=\overline{p_jp_{j+1}}$. In both cases, we have $\overline{uv}=e_1$ and $\overline{xy}=e_2$. The truth value of $\ell(\overline{pq})\cap D_{r,2}(\overline{uv},\overline{xy})\neq \emptyset$ can be determined with the help of the intersection of $\ell(\overline{pq})$ with $B_{r}(u), B_{r}(v), B_{r}(x), B_{r}(y), R_{r}(\overline{uv})$ and $R_{r}(\overline{xy})$. If and only if there is an overlap of the intersection of $\ell(\overline{pq})$ with any of these geometric objects belonging to the first stadium and the intersection of $\ell(\overline{pq})$ with any of these geometric objects belonging to the second stadium, then the predicate is true. By Lemma~\ref{lem:ball:line} and Lemma~\ref{lem:capcyl:line}, it is a simple predicate to check which of these intersections exists and it can be decided with the help of a constant number of simple predicates which candidates define each of the intersections. All candidates for intersection points have the form  
     \[v= p+t(P\cup Q)(q-p)\]
with  $t(P\cup Q)=f(P\cup Q)+h(P\cup Q)\sqrt{g(P\cup Q)}$
    where $f,g$ and $h$ are well behaved functions. So by Lemma~\ref{obs:order}, the order of two candidates along $\ell(\overline{pq})$ is decided by a simple predicate. Comparing the order of all pairs of candidates determines the order of all candidates along the line. Together with the information on which intersections exist and which candidates determine the intersections, one can decide if $\ell(\overline{pq})\cap D_{r,2}(\overline{uv},\overline{xy})\neq \emptyset$. Since this information is given by a constant number of simple predicates, the whole predicate $\Pred$ is simple.
\end{proof}

\vspace{6pt}

\begin{lemma}\label{lem:p78:eff}
For any two polygonal curves $P\in\XX^{d}_m, Q\in\XX^d_k$ and a radius $r\in\RR_+$, if we treat $P$ and $Q$ as points in $\RR^{dm}$ and $\RR^{dk}$, respectively, by concatenating their vertex coordinates, then each of the predicates of types $\Pred_{\ref{fpc1}}$, $\Pred_{\ref{fpc2}}$, $\Pred_{\ref{fpc3}}$, and $\Pred_{\ref{fpc4}}$ is a simple function that maps $(P,Q,r) \in \RR^{\dim m} \times \RR^{dk+1}$ to $\{0,1\}$.
\end{lemma}
\begin{proof}
For $\Pred_{\ref{fpc1}}$ and $\Pred_{\ref{fpc2}}$, this statement directly follows from Lemma~\ref{lem:point:edge} if we interpret points $q_1$ and $q_k$ in $\Pred_{\ref{fpc1}}$ and $\Pred_{\ref{fpc2}}$ as degenerate edges $\overline{q_1 q_1}$ and $\overline{q_k q_k}$. Let $\Pred$ be a predicate of type $\Pred_{\ref{fpc3}}$ or $\Pred_{\ref{fpc4}}$ with input $(P,Q,r)$.  The truth value of $\Pred$ can be determined by checking if a line segment $\overline{pq}$ intersects two balls $B_{r}(u)$ and $B_{r}(v)$. For $\Pred=\Pred_{\ref{fpc3}}$, we have $\overline{pq}=\overline{q_iq_{i+1}}$, $u=p_j$ and $v=p_t$. For $\Pred=\Pred_{\ref{fpc4}}$, we have $\overline{pq}=\overline{p_jp_{j+1}}$, $u=q_i$ and $v=q_t$. To answer the predicate, one can compute the intersections of the line $\ell(\overline{pq})$ with each of the balls $B_{r}(u)$ and $B_{r}(v)$ and then check if they are in the right order.
\end{proof}

\subsection{Predicates for polygonal regions that may contain holes}\label{sec:proof:Hausd:Pol}

In the following, we show that each of the predicates $\Pred_{\ref{hprb1}},\dots,\Pred_{\ref{hpri3}}$ is either simple or a combination of a polynomial number of simple predicates.

\vspace{9pt}

\begin{lemma}\label{lem:hausd:pol:eff1}
For any two polygonal regions $P \in \PP_m$ and $Q\in \PP_k$ and any positive radius $r$, if we treat $P$ and $Q$ as points in $\RR^{3m}$ and $\RR^{3k}$, respectively, by concatenating their vertex coordinates and boundary labels, each of the predicates of types $\Pred_{\ref{hprb2}}$, $\Pred_{\ref{hprb3}}$, $\Pred_{\ref{hprb4}}$, $\Pred_{\ref{hprb5}}$, and $\Pred_{\ref{hpri1}}$ is a simple function that maps $(P,Q,r) \in \RR^{3m} \times \RR^{3k+1}$ to $\{0,1\}$.
\end{lemma}
\begin{proof}
Let $\Pred$ be a predicate  with input $(P,Q,r)$.
If $\Pred$ is of type $\Pred_{\ref{hprb2}}$, then it directly follows by Lemma~\ref{lem:edge:edge} that $\Pred$ is simple.
If $\Pred$ is of type $\Pred_{\ref{hprb3}}$ then it is a simple predicate to check (Lemma~\ref{lem:edge:edge}) if the two intersections exist and as described in Lemma~\ref{obs:order}, it needs only a constant number of simple predicates to determine the order of the intersections (if they exist). 
If $\Pred$ is of type $\Pred_{\ref{hprb4}}$ or $\Pred_{\ref{hprb5}}$, then it is a simple predicate (Lemma~\ref{lem:edge:edge}) to check  if the two intersections exist and which points are the first and the last points of the intersection (if they exist). Since all candidates for the first and last point are of root-type 1, the distance of each of the candidates to the edge $e_3$ can be checked with a simple predicate by Lemma~\ref{lem:point:edge}. If $\Pred$ is of type $\Pred_{\ref{hpri1}}$ then it directly follows by Lemma~\ref{lem:point:edge} that $\Pred$ is simple because all Voronoi-vertex-candidates are vertices of root-type 1, 2 or 3 by Lemma~\ref{lem:vertex_candidates}.
\end{proof}

\vspace{6pt}

\begin{lemma}\label{lem:hausd:pol:eff2}
For any two polygonal regions $P\in \PP_m$ and $Q\in \PP_k$ and a positive radius $r$, if we treat $P$ and $Q$ as points in $\RR^{3m}$ and $\RR^{3k}$, respectively, by concatenating their vertex coordinates and boundary labels, then each of the predicates of types $\Pred_{\ref{hprb1}}$, $\Pred_{\ref{hpri2}}$, and $\Pred_{\ref{hpri3}}$ can be determined by a polynomial number (in $k$ and $m$) of simple predicates as functions that map $(P,Q,r) \in \RR^{3m} \times \RR^{3k+1}$ to $\{0,1\}$.
\end{lemma}
\begin{proof}
    Let $\Pred$ be a predicate of type $\Pred_{\ref{hprb1}},\Pred_{\ref{hpri2}}$ or $\Pred_{\ref{hpri3}}$ with input $(P,Q,r)$. The truth value of $\Pred$ can be determined by checking if a vertex $v$ is contained in a polygonal region $A\in\{P,Q\}$. In all cases, $v$ is a point of root-type 1, 2 or 3 (see Lemma~\ref{lem:vertex_candidates}). Consider the following two types of predicates.
    \begin{itemize}
\item $(\Pred')$ : Given an edge $e$ of $A$, this predicate returns true if and only if $hr(v)\cap e\neq\emptyset$.
\item $(\Pred'')$ : Given a vertex $a$ of $A$, this predicate returns true if and only if $hr(v)\cap a\neq\emptyset$.
\end{itemize}
Knowing all of these predicates can determine how many times the horizontal ray $hr(v)$ crosses the boundary of $A$. If $hr(v)$ crosses the boundary an even amount of times, then $v\notin A$ and for an odd amount of times, we have $v\in A$. The vertices have to be considered in $\Pred''$ to not count any intersection twice. Each predicate of the form $\Pred'$ or $\Pred''$ is simple by Lemma~\ref{lem:ray:edge} (interpret a vertex $a$ as a degenerate edge $\overline{aa}$). Since there are only a polynomial number of predicates of the form $\Pred'$ and $\Pred''$, we conclude that $\Pred$ can be determined by a polynomial number of simple predicates.
\end{proof}

\subsection{VC-dimension results}\label{sec:res:end}
In the previous sections, it was shown that all predicates for all analyzed range spaces of the form $\RSpace_{\rho,k}$ can be determined by a polynomial number of simple predicates. Together with Corollary~\ref{cor:vc}, this implies our following results on their VC-dimension.

\vspace{9pt}

\begin{theorem}\label{thm:result:main}
Let $\mathcal{R}_{\mathit{H},k}$ be the range space of balls centered at polygonal curves in $\XX^d_k$ with ground set $\XX^d_m$ with respect to the Hausdorff distance.  Let $\mathcal{P}_{\mathit{H},k}$ be the range space of balls centered at polygonal regions that may contain holes in $\PP_k$ with ground set $\PP_m$ with respect to the Hausdorff distance. Then, $\mathit{VCdim}(\RSpace_{\mathit{H},k}) = O(dk\log(km))$ and $\mathit{VCdim}(\mathcal{P}_{\mathit{H},k}) = O(k\log(km))$.
    
\end{theorem}

\vspace{9pt}

\begin{theorem}\label{thm:result:main:frechet}
Let $\RSpace_{\rho,k}$ be the range space of balls under distance measure $\rho$ centered at polygonal curves in $\XX^d_k$ with ground set $\XX^d_m$. 
Then, both $\mathit{VCdim}(\RSpace_{F,k})$ and $\mathit{VCdim}(\mathcal{R}_{\mathit{wF},k})$ are $O(dk\log(km))$.
    
\end{theorem}

\begin{proof}[Proof of Theorems~\ref{thm:result:main}, \ref{thm:result:main:frechet}]
    The number of predicates of each type $\Pred_{\ref{hpc1}}, \dots \Pred_{\ref{hprb5}}$ is polynomial in $k$ and $m$. By Lemma~\ref{lem:hausd:pred}, \ref{lem:frechet:pred}, \ref{lem:wfrechet:pred}, \ref{lem:hausd:pol:pred1} and \ref{lem:hausd:pol:pred2} the relevant distance queries are determined by the truth values of these predicates. Furthermore, Lemmas \ref{lem:p3478:eff}, \ref{lem:p78:eff}, \ref{lem:hausd:pol:eff1}, and \ref{lem:hausd:pol:eff2} imply that all these predicates are determined by a polynomial number (with respect to $m$ and $k$) of simple predicates. Therefore, applying Corollary~\ref{cor:vc} directly results in the claimed bounds on the VC-dimension.
\end{proof}

\subsection{Running time of distance queries}\label{sec:distq}
In the previous sections, we have shown how to determine distance queries based on predicates. In this section, we analyze the running time of this procedure. Each of the considered distance queries can be determined based on the sign values of polynomials of constant degree. Therefore, there always exists a decision tree that takes the sign values of these polynomials as input and answers the corresponding distance query in linear time with respect to the number of polynomials.  See Table~\ref{tab:distq:rt} for an overview of the number of polynomials, such that the corresponding distance query can be determined based on their sign values.

\begin{table}[]
\renewcommand{\arraystretch}{1.4}
    \centering
\begin{tabular}{|c|c|c|}
\hline
Object & $\rho$ & Number of polynomials \\
\hline \hline
\multirow{6}{*}{polygonal curves  in $\mathbb{R}^d$} & DTW &  $O(\min \{ m^{k-1} , k^{m-1} \})$  \\

& dH & {$O(km)$} \\

     & dF  &  {$O(km)$}  \\ 
& H & {$O(km(k+m))$}  \\

    & F & {$O(km(k+m))$} \\

    & wF & {$O(km)$}   \\
\hline
    polygonal regions in $\mathbb{R}^2$ & H & $O(k^4+m^4)$  \\
    \hline
    
\end{tabular}
    \caption{An overview of the number of polynomials based on which the distance queries can be determined.  All polynomials have constant degree. If the sign values of the polynomials is given as the input, the running time to answer the distance query is linear in the number of these polynomials. }
    \label{tab:distq:rt}
\end{table}

\section{Applications}
To state our results in general terms for all considered distance measures, we again write $d_\rho$ instead of $d_H, d_F, d_{dF}, d_{wF}$ or $d_{DTW}$, when discussing techniques that can be applied for all the distance measures.

\subsection{Simplification}

\new{Take a curve $P \in \XX^d_m$.  The first problem is that given $r> 0$, compute a mimimum size $k$ and an element $Q\in \XX^d_k$ such that $d_\rho(Q,P) \leq r$.   We enumerate $b$ from 2 to $m$ until we can find an element $Q = (q_1,\ldots,q_b) \in \XX^d_b$ such that $d_\rho(Q,P) \leq r$. When this happens, we obtain the desired element of the minimum size. Let $s$ be the size of the set of polynomials $\mathcal{P}$ such that $d_\rho(Q,P)\leq r$ can be determined based on their sign values. By Theorem~\ref{thm:arr}(ii), we can compute in $O(s)^{O(db)}$ time a set $R$ of points such that $R$ contains at least one point in each cell of $\mathscr{A}(\mathcal{P})$, as well as the sign condition vectors for $\mathcal{P}$ at the points in $R$. See Table~\ref{tab:distq:rt} for the values of $s$. As discussed in Section~\ref{sec:distq}, it takes another $O(s)$ time per cell to determine whether the inequality $d_\rho(Q,P) \leq r$ is satisfied by the point(s) of $R$ in that cell. 
  If the answer is yes for some cell in $\mathscr{A}(\mathcal{P})$, we stop; any point in $R$ in that cell gives the desired curve $Q$.  If the answer is no for every cell in $\mathscr{A}(\mathcal{P})$, we increment $b$ and repeat the above.  Let $k$ be the minimum size of $Q$.  The total running time is 
$O(s)^{O(dk)}$.}

\new{The second problem is that given an integer $k \geq 2$, compute an element $Q \in \XX^d_k$ that minimizes $d_\rho(Q,P)$. In this case $r$ is a variable.  By Theorem~\ref{thm:arr}(ii) and as discussed in the previous paragraph, we can determine in $O(s)^{O(dk)}$ time the subset $\mathcal{K}$ of cells of $\mathscr{A}(\mathcal{P})$ that satisfy the inequality $d_\rho(Q,P) \leq r$.  Specifically, we obtain a set $R$ of points such that every point in $R$ lies in a cell of $\mathcal{K}$, and every cell in $\mathcal{K}$ contains a point in $R$, and we also obtain the sign condition vectors at the points in $R$ which give the polynomial inequalities and equalities that describe every cell in $\mathcal{K}$.  This yields a collection of semialgebraic sets.  We invoke Theorem~\ref{thm:arr}(iii) to determine in $O(s)^{O(dk)}$ time the minimum $r$ attained in each such semialgebraic set.  The minimum over all such sets is the minimum $d_\rho(Q,P)$ desired.  The total running time is $O(s)^{O(dk)}$. }

\new{
\begin{theorem}
Take a curve $P \in \XX^d_m$, a positive value $r$, and a distance measure $d_\rho$. Let $\mathcal{P}$ be the set of polynomials such that $d_\rho(Q,P)\leq r$ can be determined based on the sign values of $\mathcal{P}$ for any $Q\in \XX^d_k$.  Let $s = |P|$.  We can compute in $O(s)^{O(dk)}$ time the element $Q$ of the minimum size $k$ that satisfies $d_\rho(Q,P) \leq r$.  For every integer $k \geq 2$, we can compute in $O(s)^{O(dk)}$ time the element $Q \in \XX^d_k$ that minimizes $d_\rho(Q,P)$. See Table~\ref{tab:distq:rt} for the values of $s$ for polygonal curves.
\end{theorem}
}

\subsection{Range searching}\label{sec: range_searching}

\new{Let $T = \{\tau_1,\ldots,\tau_n\}$ be $n$ elements of $
\XX^d_m$ or $\PP_m$.   Let $k \geq 2$ be a given integer.  We want to construct a data structure such that for any $r > 0$ and any element $Q = (q_1,\ldots,q_k) \in \XX^d_k$ or $\PP_k$ corresponding to the type of elements in $T$, we can report the subset of $T$ that are within a distance $r$ from $Q$.  Let $\mathcal{P}$ be the set of polynomials s.t. $d_\rho(Q,P)\leq r$ can be determined based on their sign values. }

\new{Every cell in $\mathscr{A}(\mathcal{P})$ represents a subset $T' \subseteq T$ such that for every $\tau_a \in T'$ and every point $(q_1,\ldots,q_k,r)$ in that cell, $d_\rho((q_1,\ldots,q_k),\tau_a) \leq r$.  Conceptually speaking, it suffices to perform a point location in $\mathscr{A}(\mathcal{P})$ using the query point $(q_1,\ldots,q_k,r)$.  This can be accomplished using a tree that represents a hierarchical decomposition; each node stores a small subset of the polynomials so that we can compare the query point with the arrangement of this small subset to decide which child to visit.  For example, the point enclosure data structure in~\cite{AAEZ2021} is organized like this.  Unfortunately, the arrangement of this small subset of polynomials at each node has size exponential in the ambient space dimension.  In our case, this dimension is $dk+1$, so querying takes time exponential in $dk+1$ at each node which is undesirable.}

\new{Fortunately, as stated in Theorem~\ref{thm:locate}, point location in an arrangement of hyperplanes has a much better dependence on the ambient space dimension.  We linearize the polynomials in $\mathcal{P}$, that is, we introduce a new variable to stand for every product of monomials of variables. Since each considered polynomial  has constant degree and involves at most a constant number of vertices of $Q$, after linearization, the total number of variables cannot be more than $O(dk)^{O(1)}$. There are $O(kmn)^{O(1)}$ polynomials in $\mathcal{P}$ for all considered distance measures (see Table~\ref{tab:distq:rt}). Hence, linearization produces a set of $O(kmn)^{O(1)}$ hyperplanes in $O(dk)^{O(1)}$ dimensions.  Building the point location data structure in Theorem~\ref{thm:locate} for this arrangement of hyperplanes solves our problem.  }

\new{\begin{theorem}
	\label{thm:range}
	Let $T = \{\tau_1,\ldots,\tau_n\}$ be a set of $n$ elements in $\XX^d_m$ or $\PP_m$.  Let $k \geq 2$ be a given integer. We can construct a data structure of $O(kmn)^{\mathrm{poly}(d,k)}$ size in $O(kmn)^{\mathrm{poly}(d,k)}$  expected time such that for $r > 0$ and any query curve $Q \in \XX^d_k$ or $\PP_k$ corresponding to the type of elements in $T$, the subset of $T$ that are within a Fr\'{e}chet distance of $r$ from $Q$ can be reported in time $O((dk)^{O(1)}\log(kmn))$ plus the output size.
\end{theorem}}

\subsection{Nearest neighbor and distance oracle}

\new{
  Let $T = \{\tau_1,\ldots,\tau_n\}$ be $n$ elements in $\XX^d_m$ or $\PP_m$.  Let $k \geq 2$ be a given integer.  We construct the set of polynomials $\mathcal{P}$ for $T$ as in Section~\ref{sec: range_searching} for range searching.  The variables are $r$ and the coordinates of the vertices of $Q = (q_1,\ldots,q_k)$.  So the ambient space is $\real^{dk+1}$.  Without loss of generality, let the $r$-axis be the vertical axis.}

\new{Let $\mathcal{K}$ be the subset of cells in $\mathscr{A}(\mathcal{P})$ that represent non-empty range searching results.  Let $\bigcup \mathcal{K}$ denote the union of cells in $\mathcal{K}$.  Observe that $\bigcup \mathcal{K}$ is \emph{upward monotone} in the sense that its intersection with any vertical line is either empty or a halfline that extends vertically upward.  It is because if $(Q,r) \in \bigcup \mathcal{K}$, there is a non-empty subset $T' \subseteq T$ such that all elements in $T'$ are within a distance of $r$ from $Q$; therefore, for any $r' > r$, all elements in $T'$ are also within a distance of $r'$ from $Q$.  The upward monotonicity of $\bigcup \mathcal{K}$ allows us to show the next result.}

\new{\begin{lemma}
	\label{lem:lower}
	The lower boundary of $\bigcup \mathcal{K}$ is $\bigcup \mathcal{L}$ for some subset $\mathcal{L} \subseteq \mathcal{K}$.   Moreover, $\mathcal{L}$ can be constructed in $O(kmn)^{O(dk)}$ time.
\end{lemma}}
\begin{proof}
\new{
It suffices to prove that a cell $C \in \mathcal{K}$ cannot lie partially in the lower boundary of $\bigcup \mathcal{K}$.  Assume to the contrary that a portion of $C$ lies in the lower boundary of $\bigcup \mathcal{K}$, but a portion of $C$ does not.  It follows that there is a point $p$ in the interior of $C$ such that $p$ lies in the lower boundary of $\bigcup \mathcal{K}$, but any arbitrarily small open neighborhood of $p$ in $C$ contains a point of $C$ that does not lie in the lower boundary of $\bigcup \mathcal{K}$.  Shoot a ray $\gamma$ vertical downward from $p$.  If $\gamma$ intersects another cell in $\mathcal{K}$, then $p$ cannot lie in the lower boundary of $\bigcup \mathcal{K}$, a contradiction.  Suppose that $\gamma$ does not intersect another cell in $\mathcal{K}$.  Then, there must exist some open neighborhood $N_p$ of $p$ in $C$ such that one can shoot vertical rays downward from points in $N_p$ without intersecting another cell in $\mathcal{K}$.  But then $N_p$ must be part of the lower boundary of $\bigcup \mathcal{K}$, a contradiction to what we said earlier about arbitrarily small open neighborhoods of $p$ in $C$.    This completes the proof of the first part of the lemma.}

\new{We construct $\mathcal{L}$ as follows.  First, by Theorem~\ref{thm:arr}(ii), we spend $O(kmn)^{O(dk)}$ time to construct a set $R$ of points that contain points in each cell in $\mathscr{A}(\mathcal{P})$.  The sign condition vectors at the points in $R$ are also computed.  Note that the cardinality of $R$ is $O(kmn)^{O(dk)}$.  For every cell $C$ in $\mathscr{A}(\mathcal{P})$, a point in $R \cap C$ represents an element a value $r$ and an element $Q\in \XX^d_k$ or $\PP_k$, and we check whether $d_\rho(Q,\tau_a) \leq r$ for all $a \in [n]$ in $O(kmn)^{O(1)}$ time, which tells us whether $C \in \mathcal{K}$.  For every cell $C \in \mathcal{K}$, we shoot a vertical ray downward from a point $p \in R \cap C$ to see if the ray intersects another cell in $\mathcal{K}$.  This check is done as follows.  Take any other cell $C' \in \mathcal{K}$.   The sign condition vector of a point in $R \cap C'$ gives the polynomial inequalities and equalities that describe $C'$.  Let $p = (q_1,\ldots,q_k,r_0)$.  Plug $q_1,\ldots,q_k$ into the polynomials in the description of $C'$.  Note that the polynomials are independent of $r$, or have degree at most $4$ in $r^2$ (i.e. have the form $\sum_{i=0}^4 c_ir^{2i}$ for constants $c_i$).  Only those that involve $r$ are of interest to us.  In $O(dkm^2n)$ time, we can solve for the conditions on $r$ imposed by the polynomials in the description of $C'$.  In general, each polynomial equality/inequality specify $O(1)$ disjoint ranges of $r$ for which the polynomial equality/inequality is satisfied.  Starting with the range $r < r_0$ imposed by $p$, we can examine each polynomial in turn to ``accumulate'' the disjoint feasible ranges of $r$.  That is, if $\mathcal{R}$ is the current list of disjoint feasible ranges of $r$, then for every range $R$ of $r$ given by the next polynomial, we  compute the new ranges $\{R \cap R' : R' \in \mathcal{R}\}$.  The cardinality of $\mathcal{R}$ increases by $O(1)$ after processing each polynomial.   Hence, we can decide in $\tilde{O}(km^2n)$ time whether the downward ray from $p$ intersects $C'$ or not, which means that we can decide in $O(kmn)^{O(dk)}$ time whether $C$ belongs to $\mathcal{L}$.  In all, we can construct $\mathcal{L}$ in $|\mathcal{L}| \cdot O(kmn)^{O(dk)} = O(kmn)^{O(dk)}$ time.}
\end{proof}

\vspace{3pt}

\new{We are interested in $\mathcal{L}$ because for any $Q \in \XX^d_k$ or $\PP_k$, if $(Q,r)$ belongs to some cell $C \in \mathcal{L}$, then $C$ must represent either exactly one element $\tau_a \in T$ as the range searching result using $(Q,r)$ or multiple elements that all have the same distance to $Q$.  It follows that each element in the range searching result is a nearest neighbor of $Q$.  Therefore, we want to perform point location in the vertical projection of the cells in $\mathcal{L}$ into $\real^{dk}$.  By Lemma~\ref{lem:proj}, the polynomials that define the downward projection of $\mathcal{L}$ have constant degree, and they can be computed in $|\mathcal{L}| \cdot O(kmn)^{O(dk)} = O(kmn)^{O(dk)}$ time.  It follows that there are  $O(kmn)^{O(dk)}$ polynomials that define the projection of $\mathcal{L}$ in $\real^{dk}$.  To support point location in the projection of $\mathcal{L}$, we linearize the polynomials in the projection to form hyperplanes in $O(dk)^{O(1)}$ dimensions.  Then, we apply Theorem~\ref{thm:locate} to these hyperplanes to obtain a point location data structure.   Doing the point location in this arrangement of hyperplanes gives the nearest neighbor of $Q$. In order to report the nearest neighbor distance, each cell $C$ in the projection of $\mathcal{L}$ corresponds to a cell $\hat{C}$ in the arrangement of the hyperplanes after linearization, so we store at $\hat{C}$ one of the polynomial equalities in the definition of the preimage of $C$ that involves $r$.  Then, after the point location, we can solve that polynomial equality for $r$.  Every polynomial equality has at most degree $4$ in $r^2$ and therefore can be solved in $O(d)$ time.}  

\new{\begin{theorem}
	\label{thm:nearest}
	Let $T = \{\tau_1,\ldots,\tau_n\}$ be $n$ elements in $\XX^d_m$ or $\PP_m$.  Let $k \geq 2$ be a given integer.  We can construct a data structure of $O(kmn)^{\mathrm{poly}(d,k)}$ size in $O(kmn)^{\mathrm{poly}(d,k)}$ expected time such that for any $Q \in \XX^d_k$ or $\PP_k$ corresponding to the type of elements in $T$, we can find its nearest neighbor in $T$ under $d_\rho$ in $O((dk)^{O(1)}\log (kmn))$ time.
\end{theorem}}

 \new{Theorem~\ref{thm:nearest} gives a distance oracle in the special case of $n = 1$.}

\new{\begin{theorem}
	Let $\tau$ be an element in $\XX^d_m$ or $\PP_m$.  Let $k \geq 2$ be a given integer.  We can construct a data structure of $O(km)^{\mathrm{poly}(d,k)}$ size in $O(km)^{\mathrm{poly}(d,k)}$ expected time such that for any $\sigma \in \mathbb{X}_k^d$ or $\PP_k$ corresponding to the type of elements in $T$, we can return $d_\rho(\sigma,\tau)$ in $O((dk)^{O(1)}\log (km))$ time.
\end{theorem}}

\section{Conclusion}

We demonstrate a connection between algebraic geometry and distance measures, including the Fr\'{e}chet distance, the Hausdorff distance and DTW, that allows us to obtain improved VC dimension bounds and exact algorithms for several fundamental problems concerning these distance measures.  When $d$ and $k$ are $O(1)$, our results imply polynomial-time simplification algorithms and data structures for range searching, nearest neighbor search, and distance determination that have polynomial space complexities and logarithmic query times.   The connection to algebraic geometry may offer new perspectives in designing new algorithms and proving approximation results.  Our approach is general enough that it should be possible to handle other variants of the problems considered.  For example, when performing range searching, one may want to return inclusion-maximal subcurves that are within a Fr\'{e}chet distance of $r$ from the query curve $Q$, although the whole curve may be more than a Fr\'{e}chet distance of $r$ away from $Q$.  It seems possible to introduce more polynomials to capture which vertices or edges of an input curve $\tau$ are matched by the endpoints of $Q$ under the  Fr\'{e}chet matching.

\vspace{9pt}

\backmatter

\bmhead{Acknowlegments}

The research of Cheng and Huang is supported by the Research Grants Council, Hong Kong, China (project no.~16208923). The research of Brüning and Driemel is supported by the Deutsche
Forschungsgemeinschaft (DFG, German Research Foundation, project no. 390685813; 459420781). Anne Driemel is affiliated with the Lamarr Institute for Machine Learning and Artificial Intelligence lamarr-institute.org.

\bibliographystyle{plain}
\bibliography{biblio}

\end{document}